\documentclass[onecolumn,final]{svjour2}

\usepackage{graphicx}
\usepackage{latexsym}
\usepackage{revsymb}
\usepackage{amsfonts}
\usepackage{amsmath}
\usepackage{amssymb}
\usepackage{bm}

\journalname{Foundations of Physics}

\newcommand{\bra}[1]{\langle #1 |}
\newcommand{\ket}[1]{| #1 \rangle}
\newcommand{\braket}[2]{\langle #1 | #2\rangle}
\newcommand{\inner}[2]{\bigl(#1,#2\bigr)}
\newcommand{\dotproduct}[2]{#1\!\cdot\!#2}
\newcommand{\ppm}{\varphi}
\newcommand{\ppmf}[1]{\ppm\left(#1\right)}

\newcommand{\V}[2]{V(#1,#2)}
\newcommand{\R}{\mathbb{R}}
\newcommand{\C}{\mathbb{C}}
\newcommand{\Z}{\mathbb{Z}}
\newcommand{\F}{\mathbb{Z}}
\newcommand{\Fp}{\mathbb{Z}_p}
\newcommand{\Fpi}{\mathbb{Z}_p[\,\underline{i}\,]}
\newcommand{\Fthreei}{\mathbb{Z}_3[\,\underline{i}\,]}
\newcommand{\GFsigma}{\hat{\sigma}}
\newcommand{\Csigma}{\tilde{\sigma}}

\newcommand{\0}{\underline{0}}
\newcommand{\1}{\underline{1}}
\newcommand{\2}{\underline{2}}

\renewcommand{\i}{\underline{i}}
\renewcommand{\a}{\underline{a}}
\renewcommand{\b}{\underline{b}}
\renewcommand{\c}{\underline{c}}
\renewcommand{\d}{\underline{d}}
\newcommand{\g}{\underline{g}}
\newcommand{\x}{\underline{x}}

\begin{document}

\title{Biorthogonal Quantum Mechanics:\\
Super-Quantum Correlations and Expectation Values without Definite Probabilities
}
\titlerunning{Biorthogonal Quantum Mechanics}

\author{Lay Nam Chang
\and
Zachary Lewis
\and
Djordje Minic
\and 
Tatsu Takeuchi
}

\institute{
L. N. Chang, Z. Lewis, D. Minic, T. Takeuchi\\
Department of Physics, Virginia Tech, Blacksburg, VA 24061, USA\\
\email{laynam@vt.edu, zlewis@vt.edu, dminic@vt.edu, takeuchi@vt.edu}\\
}

\date{August 26, 2012}

\maketitle

\begin{abstract}
We propose mutant versions of quantum mechanics constructed on vector spaces
over the finite Galois fields $GF(3)$ and $GF(9)$. 
The mutation we consider here is distinct from what we proposed in previous papers
on Galois field quantum mechanics. 
In this new mutation, the canonical expression for expectation values is
retained instead of that for probabilities. 
In fact, probabilities are indeterminate.
Furthermore, it is shown that the mutant quantum mechanics over 
the finite field $GF(9)$ exhibits super-quantum correlations 
(i.e. the Bell-Clauser-Horne-Shimony-Holt bound is 4).
We comment on the fundamental physical importance of these results in the context of quantum gravity.
\end{abstract}


\section{Introduction}

Quantum descriptions of physical systems
begin with the introduction of a vector space, generally defined over the complex number field $\C$, 
with elements of the space associated with states of the physical system under consideration. 
This space, in the traditional approach, is assumed to be a Hilbert space $\mathcal{H}$, 
which for $N$-level systems is $\mathcal{H}=\C^N$.
The Hilbert space $\mathcal{H}$ possesses a natural inner product 
$\mathcal{H}\times\mathcal{H}\rightarrow\C$,
which we denote
\begin{equation}
\inner{\ket{\alpha}}{\ket{\beta}}\,\in\,\C\;,\qquad
\ket{\alpha},\,\ket{\beta}\,\in\,\mathcal{H}\;.
\end{equation}
It is customary to associate a dual-vector $\bra{\alpha}\in\mathcal{H}^*$ to each vector $\ket{\alpha}\in\mathcal{H}$,
with the same label $\alpha$, via
\begin{equation}
\bra{\alpha}\;=\;\inner{\ket{\alpha}}{\phantom{\ket{\beta}}}\;,
\end{equation}
so that
\begin{equation}
\braket{\alpha}{\beta}\;=\;\inner{\ket{\alpha}}{\ket{\beta}}\;.
\end{equation}
The presence of the inner product allows the definition
of hermitian conjugation of linear operators via
\begin{equation}
\inner{\ket{\alpha}}{\hat{A}\,\ket{\beta}}\;=\;
\inner{\hat{A}^\dagger\ket{\alpha}}{\ket{\beta}}\;,
\end{equation}
and that of hermitian operators, to which physical observables are associated,
via $\hat{A}^\dagger=\hat{A}$.
It also allows for the definition of unitary operators via
\begin{equation}
\inner{\hat{U}\ket{\alpha}}{\hat{U}\ket{\beta}}
\;=\;
\inner{\ket{\alpha}}{\ket{\beta}}
\;,
\end{equation}
under which the states are assumed to evolve.

There are two ways for the quantum description to make contact with physical reality.    
In the first approach, possible outcomes of a measurement of an observable $\hat{A}$
are assumed to be given by its eigenvalues.
Let us denote the eigenvector associated with eigenvalue $\alpha$ by $\ket{\alpha}$:
\begin{equation}
(\hat{A}-\alpha)\ket{\alpha}\;=\;0\;.
\end{equation}
When the system is in the state represented by $\ket{\psi}\in\mathcal{H}$,
the probability of obtaining the outcome $\alpha\in\R$ as a result of a measurement of $\hat{A}$ is given by 
\begin{equation}
P(\alpha|\psi)\;=\;
\dfrac{\bigl|\braket{\alpha}{\psi}\bigr|^2}{\sum_\beta \bigl|\braket{\beta}{\psi}\bigr|^2}
\;,
\label{ProbDef}
\end{equation}
where the sum in the denominator runs over all the eigenvalues of $\hat{A}$.
The hermiticity of $\hat{A}$ ensures that its eigenvalues are all real, and
that the eigenvectors are mutually orthogonal and complete.
Normalizing the state vector and the eigenvectors of $\hat{A}$ so that
\begin{equation}
\braket{\psi}{\psi}=1\;,\qquad\braket{\alpha}{\beta}\;=\;\delta_{\alpha\beta}\;,
\end{equation}
the above expression for the probability reduces to 
$P(\alpha|\psi)=|\braket{\alpha}{\psi}|^2$.

There is an alternative way of making contact with reality, which is equivalent to the one above for conventional treatments.    
One begins with the quantity
\begin{equation}
\dfrac{\bra{\psi}\hat{A}\ket{\psi}}{\braket{\psi}{\psi}}\;,
\label{EVdef}
\end{equation}
which is real for hermitan $\hat{A}$, and interprets the result as the expectation value for the associated observable in the state $\ket{\psi}$.   
If $\braket{\psi}{\psi}=1$, the expression reduces to 
$\bra{\psi}\hat{A}\ket{\psi}$.
Note that this is the standard approach in Quantum Field Theory (QFT) where all physical predictions are
expressed in terms of $N$-point correlation functions, \textit{i.e.}, the vacuum expectation values of the products
of $N$ field operators. 
This is most explicit in the path integral formulation of QFT.
To recover the probabilistic interpretation of the first approach, 
one asserts that the probability for obtaining the outcome $\alpha$ 
for the measurement of $\hat{A}$ on the state $|\psi\rangle$ is given by 
\begin{equation}
P(\alpha|\psi) 
\;=\; \dfrac{\bra{\psi}\, \delta (\hat{A} - \alpha)\,\ket{\psi}}{\braket{\psi}{\psi}}
\;.
\label{Pdelta}
\end{equation}
No absolute values are invoked, and attention is shifted to moments of the relevant observable operator in the state in question; in particular we do need the expectation values of powers of the operator.  
For canonical quantum descriptions using the Hilbert space $\mathcal{H}$, these two starting points lead to identical results.

The situation however changes when the underlying space is not a Hilbert space.   
Indeed, for spaces for which the inner product is ill-defined, one can expect different outcomes for these two approaches.    

In Refs.~\cite{Chang:2012eh} and \cite{Chang:2012gg} (inspired by \cite{Chang:2011yt},
\cite{MQT}, and \cite{nambu}),
we have explored the possibility of discretizing the fields over which the vector space is defined
but retaining the physical interpretation provided by the first approach, namely, the 
definition of probabilities via Eq.~(\ref{ProbDef}). 
The fields we considered were finite Galois fields $GF(p^n)$, 
where $n\in\mathbb{N}$ and $p$ is a prime number.
For the $n=1$ case, they are $GF(p)=\Fp=\Z/p\Z$.
Vector spaces over $GF(p^n)$ do not have inner products
since $GF(p^n)$ is not an ordered field\footnote{Ordered fields are fields on which an ordering can be imposed that respects both addition and multiplication.},
preventing any bilinear map to $GF(p^n)$ from being positive-definite (or non-negative) 
in a natural way.

However, it was recognized that for Eq.~(\ref{ProbDef}) to make sense,
the dual-vectors that appear in the expression only need to constitute a basis for the dual-vector space with a possible outcome of a measurement associated with each one.
The usual pairing of dual-vectors with vectors via the inner product is 
inessential.  Indeed, all the inner product does, in a sense, is 
connect the two approaches via the property
\begin{equation}
\braket{\psi}{\alpha}
\;=\;\inner{\ket{\psi}}{\ket{\alpha}}
\;=\;\inner{\ket{\alpha}}{\ket{\psi}}^*
\;=\;\braket{\alpha}{\psi}^*\;,
\end{equation}
so that we can write,
\begin{eqnarray}
\sum_\alpha \alpha P(\alpha|\psi)
& = & \dfrac{\sum_\alpha \alpha\bigl|\braket{\alpha}{\psi}\bigr|^2}
            {\sum_\beta  \bigl|\braket{\beta}{\psi}\bigr|^2} 
\;=\; \dfrac{\sum_\alpha \braket{\alpha}{\psi}^* \alpha\braket{\alpha}{\psi}}
            {\sum_\beta  \braket{\beta}{\psi}^*  \braket{\beta}{\psi}} \cr
& = & \dfrac{\sum_\alpha \braket{\psi}{\alpha}\alpha\braket{\alpha}{\psi}}
            {\sum_\beta  \braket{\psi}{\beta} \braket{\beta}{\psi}} 
\;=\; \dfrac{\bra{\psi}\hat{A}\ket{\psi}}{\braket{\psi}{\psi}}\;,
\end{eqnarray}
where we have made the identification
\begin{equation}
\hat{A}\;=\;\sum_\alpha\ket{\alpha}\alpha\bra{\alpha}\;.
\label{Aspect}
\end{equation}
Thus, for the first approach, inner products are not necessary,
and once a basis of the dual-vector space
and the associated set of outcomes is specified, 
we have an `observable.'

To make contact with the outcome of measurements and probability distributions, we need a map from the Galois field to that of non-negative reals.  It is essential that this map preserves products, which is necessary to distinguish entangled states from product ones, and also for the
actions of symmetry groups on the Galois field.   This is achieved in \cite{Chang:2012eh} and \cite{Chang:2012gg} through an absolute value function.  
Eq.~(\ref{ProbDef}) can be used as is to define the probability
of each outcome via the absolute value function from $GF(p^n)$ to $\R$ 
given by
\begin{equation}
|\,\underline{k}\,|\;=\;
\begin{cases}
\;0 \quad &\mbox{if $\underline{k}=\0$}\;, \\
\;1 \quad &\mbox{if $\underline{k}\neq\0$}\;.
\end{cases}
\label{ABSdef}
\end{equation}
Here, numbers and symbols with underlines are used to denote elements of $GF(p^n)$, to
distinguish them from elements of $\R$.
Note that this function is product preserving, i.e. 
$|\underline{k}\underline{\ell}|=|\underline{k}||\underline{\ell}|$,
which is essential for
probabilities of product states to factorize.
Applying this formalism to 2-level systems,
we constructed spin-like observables for which the measurement outcomes were
$\pm 1\in\R$, and calculated the 
Clauser-Horne-Shimony-Holt (CHSH) \cite{Clauser:1969ny}
(see also \cite{bell,GHZ,Hardy:1993zza,cirelson})
bound for the model
and found that it was two, despite the fact that
no hidden variable mimic could reproduce the model's predictions.
For details, see Refs.~\cite{Chang:2012eh} and \cite{Chang:2012gg}.

In this paper, we explore
consequences of starting with the second approach to interpretation,
namely, the definition of expectation values via Eq.~(\ref{EVdef}).
Again, we consider vector spaces over the finite Galois field $GF(p^n)$, which do not have inner products.
Thus, the concepts of normalizability of states, hermiticity of operators, and 
a dual-vector as a hermitian conjugate of a vector,
must all be reexamined before we can apply Eq.~(\ref{EVdef}).
Furthermore, working in a vector space over $GF(p^n)$, the expression 
$\bra{\psi}\hat{A}\ket{\psi}=\mbox{(row vector)$\cdot$(matrix)$\cdot$(column vector)}$
will generically lead to an element of $GF(p^n)$, 
which must be mapped to an element of $\R$ 
if the result is to represent the expectation value of a measurement
of a physical observable.
While we obtain results similar to our earlier ones for certain fields, we discover significant differences in others.

In the following, we will address these points one by one
and define a `mutant' QM on vector spaces over the fields
$GF(3)=\F_3$ and then $GF(9)=\Fthreei$, where
$\underline{i}$ is the solution to the equation $\x^2+\1\;=\;\0$,
which is irreducible in $GF(3)$. 
In both cases, we will find that $\bra{\psi}\hat{A}\ket{\psi}\in GF(3)$ by construction,
which will be mapped to a number in $\R$.
Because we are looking at the expectation values of observables, the range of this map need not be restricted to the non-negative reals as in the case of the
absolute value function.
In Appendix~B we show that the requirement that this map preserve
products and actions of symmetry groups determines the map uniquely.   
It is the use of this map for specific expectation values,
instead of the absolute value function on brackets, that distinguishes 
between the two approaches to interpretation.   
We will show below that the connection to probabilities given by Eq.~(\ref{Pdelta}) 
for canonical QM is no longer valid.  
In fact, individual probability distributions are not fixed in our approach, giving rise to indeterminacies beyond those of canonical QM.   
Our earlier result in Refs.~\cite{Chang:2012eh} and \cite{Chang:2012gg} that the CHSH bound for spin-like systems over Galois fields cannot be larger than 2 was predicated upon using the first approach starting with Eq.~(\ref{ProbDef}).  
We will find that in the second approach, the CHSH bound for the $GF(3)$ case is also 2. 
For the $GF(9)$ case, however, the CHSH bound is 4, the maximum possible value.
As far as we are aware of, this is one of the first explicit examples of a non-trivial
super-quantum theory.

Before we proceed to the heart of the matter, we note that consideration of discrete mathematical structures is not only relevant from an academic point of view.
We note that such considerations have been seriously undertaken in various approaches
to the quantum structure of space and time, i.e. in various forms of quantum gravity.
The more complete literature can be found in \cite{discrete}.

The outline of the paper is as follows: 
In section II we introduce what we call biorthogonal quantum mechanics,
and in section III we present a few examples of this construction. 
Then in section IV we consider the CHCH bound and find an explicit example of a super-quantum theory. In section V we show
that in such a theory probabilities are indeterminate. We close in section VI with detailed comments about the physical relevance of our results. Various details not covered in the main
text are presented in two appendices.

\section{Biorthogonal Quantum Mechanics}

In order to adopt the definition of
expectation values via Eq.~(\ref{EVdef}) onto a vector space over the 
Galois field $GF(p^n)$, one must define the analogue of hermitian conjugation of
vectors and linear operators without reference to an inner product.
In this section, we demonstrate that this can be accomplished
via \textit{biorthogonal systems} \cite{bi}.\footnote{%
Biorthogonal systems have been discussed in Ref.~\cite{BQS}
in the context of PT Symmetric Quantum Mechanics \cite{PT}.
}

In the following,
we restrict our attention to the Galois fields $GF(p^n)$
with $p=3\,\mathrm{mod}\,4$ and $n=1$ or $2$.
As we will see below, this restriction allows 
our formalism to maintain a close parallel to quantum mechanics
defined on vector spaces over $\R$ ($n=1$ case) or $\C$ ($n=2$ case).

\subsection{Biorthogonal Systems}

As in the previous section, elements of the finite Galois field $GF(p^n)$ are 
denoted by underlined
symbols and numbers to distinguish them from elements of $\R$ or $\C$.
The $N$-dimensional vector space over $GF(p^n)$ is denoted $\V{N}{p^n}$.
A biorthogonal system is a set consisting of a basis 
$\{\ket{1},\ket{2},\cdots,\ket{N}\}$ of the vector space $\V{N}{p^n}$,
and a basis $\{\bra{1},\bra{2},\cdots,\bra{N}\}$ of the dual vector space $\V{N}{p^n}^*$ 
such that
\begin{equation}
\braket{r}{s} 
\,=\, \underline{\delta}_{rs}\;,\qquad
r,s=1,2,\cdots,N,
\end{equation}
where
\begin{equation}
\underline{\delta}_{rs}
\;=\;
\begin{cases}
\;\0\quad & \mbox{if $r\neq s$}\;, \\
\;\1\quad & \mbox{if $r=s$}\;.
\end{cases}
\end{equation}
Such a system can be constructed as follows.

\subsection{Dot Product}

First, 
denoting the $k$-th element of the vector $\ket{a}\in\V{N}{p^n}$
as $\a_k\in GF(p^n)$, 
define the `dot product' in $\V{N}{p^n}$ as
\begin{equation}
\dotproduct{\ket{a}}{\ket{b}}\;=\;
\sum_{k=1}^{N}\,
\a_k^p\,
\b_k^{\phantom{p}}
\;\in\;GF(p^n)\;.
\end{equation}
Raising an element to the $p$-th power is semilinear in $GF(p^n)$
since
\begin{equation}
(\a+\b)^p \;=\; (\a^p+\b^p)
\end{equation}
in a field of characteristic $p$.
When $n=1$, it is an identity transformation due to
Fermat's little theorem
\begin{equation}
a^{p-1}\;=\;1\;\mathrm{mod}\;p\;,\quad\forall a\in\Z\;.
\end{equation}
For the case $n=2$, $p=3\;\mathrm{mod}\,4$, it is an analogue of
complex conjugation in $\C$.
To see this, first note that the equation
\begin{equation}
\x^2+\1\;=\;\0
\end{equation}
is irreducible in $GF(p)=\Fp$ if $p=3\,\mathrm{mod}\,4$.\footnote{%
$\x^2+\1=\0$ is reducible for $p=2$ or $p=1\,\mathrm{mod}\,4$ since in
those cases $\underline{p-1}$ will be a solution.}
Denote the solutions to this equation as $\pm\i$.
Adjoining $\i$ to $GF(p)=\Fp$ gives us
$GF(p^2)=\Fpi$. Elements of this field can be expressed
as $\a+\i\,\b$, where $\a,\b\in\Fp$.
Then
\begin{equation}
(\a+\i\,\b)^p
\,=\, \a^p + \i^{p}\b^p
\,=\, \a-\i\,\b\;.
\end{equation}
Furthermore,
\begin{eqnarray}
(\a+\i\,\b)^p(\c+\i\,\d)
& = & (\a\c+\b\d)+\i(\a\d-\b\c)\;,\cr
(\c+\i\,\d)^p(\a+\i\,\b)
& = & (\a\c+\b\d)-\i(\a\d-\b\c)\;,
\end{eqnarray}
in particular,
\begin{equation}
(\a+\i\,\b)^p(\a+\i\,\b)
\;=\; \a^2+\b^2
\;\in\; \Fp\;.
\end{equation}
Therefore, 
$\dotproduct{\ket{a}}{\ket{b}}$ and
$\dotproduct{\ket{b}}{\ket{a}}$ are `complex conjugates' 
of each other, while
$\dotproduct{\ket{a}}{\ket{a}}$
is `real.'
Thus, when $p=3\,\mathrm{mod}\,4$,  
the fields $GF(p)=\Fp$ and $GF(p^2)=\Fpi$ take on the roles of
$\R$ and $\C$.

In the following, when we say $GF(p^n)$, we will mean either 
$GF(p)$ or $GF(p^2)$ with $p=3\,\mathrm{mod}\,4$
unless stated otherwise.
Also, borrowing from standard terminology,
we will say that two vectors in $\V{N}{p^n}$ are `orthogonal' to each other
when they have a zero dot product,
and that a vector is `self-orthogonal' when it is orthogonal to itself.

\subsection{Conjugation of Vectors}

Next, choose a basis $\{\ket{1},\ket{2},\cdots,\ket{N}\}$ for $\V{N}{p^n}$  
such that:
\begin{equation}
\dotproduct{\ket{r}}{\ket{s}}\;
\begin{cases}
\;\neq\; \0\qquad & \mbox{if $r=s$}\;, \\
\;=\;    \0\qquad & \mbox{if $r\neq s$}\;,
\end{cases}
\label{OrthoBasis}
\end{equation}
that is, all the basis vectors are orthogonal to each other,
but none are self-orthogonal.
Let us call such a basis an `ortho-nondegenerate' basis.
The simplest example of an ortho-nondegenerate basis would be such
that the $r$-th element of the $s$-th vector is given by
$\underline{\delta}_{rs}$,
proving that such a basis always exists.
On the other hand, not all bases satisfy this condition 
since $\V{N}{p^n}$ typically has multiple 
self-orthogonal vectors other than the zero vector.

Define the `conjugate' dual vector for each vector $\ket{r}$ 
in the ortho-nondegenerate basis as
\begin{equation}
\bra{r} \;\equiv\; 
\dfrac{\ket{r}\,\cdot}{\dotproduct{\ket{r}}{\ket{r}}}
\label{Vconjugation}
\end{equation}
where it is crucial that $\dotproduct{\ket{r}}{\ket{r}}\neq\0$
for $\bra{r}$ to exist.
Then, the set of dual vectors $\{\bra{1},\bra{2},\cdots,\bra{N}\}$
provides a basis for the dual vector space $\V{N}{p^n}^*$ 
such that $\braket{r}{s}=\underline{\delta}_{rs}$.
Thus, we obtain the set
\begin{equation}
\bigl\{ \{\bra{1},\bra{2},\cdots,\bra{N}\},\{\ket{1},\ket{2},\cdots,\ket{N}\} \bigr\}
\end{equation}
which constitutes a biorthogonal system.

\subsection{Observables}

Given a biorthogonal system, we can define the analog of
hermitian operators via
\begin{equation}
\hat{A}\;=\; \sum_{k=1}^N \underline{\alpha}_k\,\ket{k}\bra{k}\;,\qquad 
\underline{\alpha}_k\in\,GF(p)\;.
\end{equation}
Due to the biorthogonality of the system, 
$\ket{k}$ is the eigenvector of $\hat{A}$ with eigenvalue $\underline{\alpha}_k$.
Note that the eigenvalues $\underline{\alpha}_k$ are chosen to be 
elements of $GF(p)$, not $GF(p^2)$, i.e. they are `real.'

In the defining biorthogonal system, the matrix representation of
$\hat{A}$ is diagonal.
In a different biorthogonal system, say
$\bigl\{ \{\bra{1'},\bra{2'},\cdots,\bra{N'}\},\{\ket{1'},\ket{2'},\cdots,\ket{N'}\} \bigr\}$,
its matrix representation is
\begin{eqnarray}
\bra{r'}\hat{A}\ket{s'}
& = & \sum_{k=1}^N \underline{\alpha}_k\,\braket{r'}{k}\,\braket{k}{s'} \cr
& = & \sum_{k=1}^N \underline{\alpha}_k\,
\dfrac{(\dotproduct{\ket{r'}}{\ket{k}})\,(\dotproduct{\ket{k}}{\ket{s'}})}
      {(\dotproduct{\ket{r'}}{\ket{r'}})\,(\dotproduct{\ket{k}}{\ket{k}})}
\;,
\end{eqnarray}
which in general is not a hermitian matrix.
However, the diagonal elements $\bra{r'}\hat{A}\ket{r'}$
are nevertheless `real'
since $\dotproduct{\ket{r'}}{\ket{k}}$ and $\dotproduct{\ket{k}}{\ket{r'}}$
are `complex conjugates' of each other,
while $\dotproduct{\ket{r'}}{\ket{r'}}$ and $\dotproduct{\ket{k}}{\ket{k}}$
are `real.'
We identify these pseudo-hermitian operators with physical observables.

\subsection{Physical States}

Since we wish to use Eq.~(\ref{EVdef}) to define the
expectation value for the observable $\hat{A}$, 
every physical state $\ket{\psi}$ must have a conjugate dual $\bra{\psi}$,
which we define via Eq.~(\ref{Vconjugation}).
Thus, we demand that all physical states belong to some biorthogonal system.
Essentially, all vectors that are not self-orthogonal belong to
some biorthogonal system, so this requirement is equivalent to
dropping all vectors that are self-orthogonal from the 
set of physical states.

Note that if we multiply $\ket{\psi}$ with a scalar, that is,
a non-zero element of $GF(p^n)$, then its conjugate $\bra{\psi}$
will be multiplied by the inverse of that scalar.
This will leave $\hat{A}$ and $\bra{\psi}\hat{A}\ket{\psi}$ invariant.
Thus, we can identify all vectors that differ with each other
by a multiplicative scalar as representing the same 
physical state, that is, all non-zero elements of
$GF(p^n)$ can be considered to be `phases.'
For $\V{N}{p^n}$, this means that the set of physical states is 
the non-self-orthogonal subset of the projective space 
\begin{equation}
PG(N-1,p^n)\;=\;
\Bigl[\V{N}{p^n}\backslash\{\mathbf{\0}\}\Bigr]
\Big/
\Bigl[GF(p^n)\backslash\{\0\}\Bigr]\;.
\end{equation}

\subsection{Expectation Values}

With the above definitions of observables and physical states,
we can now calculate the quantity
$\bra{\psi}\hat{A}\ket{\psi}\in GF(p)=\Fp$ for observable $\hat{A}$ and state $\ket{\psi}$.
We would like to interpret this quantity as the expectation value of the
observable $\hat{A}$. 
However, if $\hat{A}$ is to represent a physical quantity such as spin,
one must map the resulting number in $GF(p)$ to a number in $\R$.

We demand that this map from $GF(p)$ to $\R$
be product preserving for reasons that will become
clear in the following.
It is easy to see that the
absolute value function given in Eq.~(\ref{ABSdef})
is a product preserving map for any $p$.
For the $p=3\,\mathrm{mod}\,4$ case, however, in addition to the
absolute value function, there is another product preserving map
which can be constructed as follows.
First, denote the generator
of the multiplicative group $GF(p)\backslash\{\0\}$ by $\g$
and express the non-zero elements of $GF(p)$ as 
$\{\g,\g^2,\g^3,\cdots,\g^{p-1}=\1\}$.
Define:
\begin{equation}
\ppmf{\x}
\;=\;
\begin{cases}
\phantom{-}0\quad & \mbox{if $\x=\0$}\;, \\
+1\quad & \mbox{if $\x=\g^\mathrm{even}$}\;, \\
-1\quad & \mbox{if $\x=\g^\mathrm{odd}$}\;.
\end{cases}
\label{phidef}
\end{equation}
It is straightforward to show that $\ppmf{\a\b}=\ppmf{\a}\ppmf{\b}$.

Note that $p=3\,\mathrm{mod}\,4$ implies
$(p-1)=\mathrm{even}$ and $(p-1)/2=\mathrm{odd}$.
Therefore,
\begin{eqnarray}
\ppmf{+\1} & = & \ppm\hspace{1px}\bigl({\g^{p-1}}\bigr) \;=\; +1 \;,\cr
\ppmf{-\1} & = & \ppm\hspace{1px}\bigl({\g^{(p-1)/2}}\bigr) \;=\; -1 \;,
\end{eqnarray}
where $-\1$ denotes the additive inverse of $\1$ in $GF(p)$.
That is, this function respectively maps $-\1$, $\0$, and $\1$ in $GF(p)$ to 
$-1$, $0$, and $1$ in $\R$.

We will use this map to give meaning to Eq.~(\ref{EVdef}) as an expectation value in the new version of quantum mechanics:
\begin{equation}
E(A|\psi)
\;=\;
\ppmf{\bra{\psi}\hat{A}\ket{\psi}}
\;.
\end{equation}
Using this identification as a starting point in modifying ordinary quantum mechanics is a viable alternative to specifying a prescription for calculating individual probabilities for outcomes of measurements, as we mentioned earlier.   The uniqueness of this map is demonstrated in Appendix B.
The rule allows us to calculate single and joint probability distributions over ensembles.


An immediate consequence of this rule is noteworthy.  The uncertainty in the measurement of $\hat{A}$
will be given by
\begin{eqnarray}
(\Delta A)^2 
& = & E(A^2|\psi)-\bigl[E(A|\psi)\bigr]^2 \cr
& = & 
\ppmf{\bra{\psi}\hat{A}^2\ket{\psi}}
-\left[\ppmf{\bra{\psi}\hat{A}\ket{\psi}}\right]^2
\;.
\end{eqnarray}
When $\ket{\psi}$ is an eigenvector of $\hat{A}$
with eigenvalue $\underline{\alpha}$,
we find
\begin{equation}
(\Delta A)^2 
\;=\; \ppmf{\underline{\alpha}^2}
-\bigl[\ppmf{\underline{\alpha}}\bigr]^2 
\;=\;0\;,
\end{equation}
due to the fact that $\ppm$ is a product
preserving map.
Thus, if a measurement of an observable is
performed on one of its eigenstates, the outcome will
always be the $\ppm$-map of the eigenvalue associated
with that state.
If $\ppm$ were not product preserving, this property would
not have been maintained.

\section{Examples}

Let us now look at a few concrete examples.

\subsection{2D Vector Space over $\bm{GF(3)}$}

Consider the 2D vector space $\V{2}{3}$
over $GF(3)=\F_3=\Z/3\Z=\{\0,\1,-\1\}$, where we denote the
additive inverse of $\1$ as $-\1$ instead of $\2$.
There are $3^2-1=8$ non-zero vectors in this space which are
\begin{equation}
\ket{a}=\left[\begin{array}{c} \1 \\ \0 \end{array}\right],\quad
\ket{b}=\left[\begin{array}{c} \0 \\ \1 \end{array}\right],\quad
\ket{c}=\left[\begin{array}{c} \1 \\ \1 \end{array}\right],\quad
\ket{d}=\left[\begin{array}{r} \1 \\ -\1 \end{array}\right],
\label{abcd}
\end{equation}
and their multiples by the `phase' $-\1$.
We find:
\begin{equation}
\begin{array}{lll}
\dotproduct{\ket{a}}{\ket{a}}& =\;\dotproduct{\ket{b}}{\ket{b}}& =\;\phantom{-}\1\;,\\
\dotproduct{\ket{c}}{\ket{c}}& =\;\dotproduct{\ket{d}}{\ket{d}}& =\;-\1\;.
\end{array}
\end{equation}
Thus, none of the vectors are self-orthogonal,
and their conjugates are
\begin{equation}
\begin{array}{llll}
\bra{a} & =\;\left[\begin{array}{cc} \1 & \0 \end{array}\right],
\quad &
\bra{c} & =\;\left[\begin{array}{cc} -\1 & \hspace{-3px}-\1 \end{array}\right], 
\\
\bra{b} & =\;\left[\begin{array}{cc} \0 & \1 \end{array}\right],
\quad &
\bra{d} & =\;\left[\begin{array}{cc} -\1 & \hspace{5px}\1 \end{array}\right].
\end{array}
\end{equation}
There are two biorthogonal systems in $\V{2}{3}^*\times\V{2}{3}$, namely
\begin{equation}
\bigl\{\{\bra{a},\bra{b}\},\,\{\ket{a},\ket{b}\}\bigr\}
\;\;\mbox{and}\;\;
\bigl\{\{\bra{c},\bra{d}\},\,\{\ket{c},\ket{d}\}\bigr\}
\;,
\label{Z3bio}
\end{equation}
up to different orderings of the vectors and dual-vectors, 
and signs.
All four inequivalent vectors belong to one of these biorthogonal systems
so they all represent physical states.

\begin{table}[t]
\centering
\begin{tabular}{|c||r|c||r|c|}
\hline
& $\quad\;\;\sigma_1\;\quad$ 
& $\quad\Delta\sigma_1\quad$ 
& $\quad\;\;\sigma_3\;\quad$ 
& $\quad\Delta\sigma_3\quad$ 
\\
\hline\hline
$\quad\ket{a}\quad$ &  $0\;\;\quad$ & $1$ &  $1\;\;\quad$ & $0$ \\
\hline
$\quad\ket{b}\quad$ &  $0\;\;\quad$ & $1$ & $-1\;\;\quad$ & $0$ \\
\hline
$\quad\ket{c}\quad$ &  $1\;\;\quad$ & $0$ &  $0\;\;\quad$ & $1$ \\
\hline
$\quad\ket{d}\quad$ & $-1\;\;\quad$ & $0$ &  $0\;\;\quad$ & $1$ \\
\hline
\end{tabular}
\caption{Expectation values and uncertainties of spin-like observables
in biorthogonal quantum mechanics on $\V{2}{3}$.}
\label{V32sigEVs}
\end{table}

We can now construct spin-like observables
with eigenvalues $\pm\1$. 
Since $\V{2}{3}$ has only two biorthogonal systems, the
two possible observables are
\begin{equation}
\begin{array}{llll}
\1\;\ket{a}\bra{a} \hspace{-2px}& -\1\;\ket{b}\bra{b} 
& = \;
\left[\begin{array}{rr} \1 & \0 \\ \0 & -\1 \end{array}\right]
& \equiv \;\GFsigma_3\;,
\\
\1\;\ket{c}\bra{c} \hspace{-2px}& -\1\;\ket{d}\bra{d} 
& = \;
\left[\begin{array}{rr} \0 & \1 \\ \1 & \phantom{-}\0 \end{array}\right]
& \equiv \;\GFsigma_1\;,
\end{array}
\label{Z3obs}
\end{equation}
up to signs. 
By construction,
$\ket{a}$ and $\ket{b}$ are respectively eigenvectors of $\GFsigma_3$ with
eigenvalues $\pm\1$.
Thus, a measurement of $\GFsigma_3$ on $\ket{a}$ will
always yield $+1$, while that on $\ket{b}$ will
always yield $-1$.
Similarly, 
$\ket{c}$ and $\ket{d}$ are respectively eigenvectors of $\GFsigma_1$ with
eigenvalues $\pm\1$, so a measurement of $\GFsigma_1$ on $\ket{c}$ will
always yield $+1$, while that on $\ket{d}$ will
always yield $-1$.

On the other hand, the expectation values of $\GFsigma_1$ and $\GFsigma_1^2$ 
for the state $\ket{a}$ are
\begin{eqnarray}
E(\sigma_1  |a)
& = & \ppmf{\bra{a}\GFsigma_1  \ket{a}}
\;=\; \ppmf{\0} \;=\; 0\;,\cr
E(\sigma_1^2|a)
& = & \ppmf{\bra{a}\GFsigma_1^2\ket{a}}
\;=\; \ppmf{\1} \;=\; 1\;,
\end{eqnarray}
so
\begin{equation}
\bigl[\Delta\sigma_1(a)\bigr]^2 \;=\;
E(\sigma_1^2|a)
-\bigl[E(\sigma_1|a)\bigr]^2
\;=\; 1\;.
\end{equation}
From these expectation values, we can infer the probabilities
of obtaining the outcomes $\pm 1$ when $\GFsigma_1$ is measured on $\ket{a}$.
Denoting these probabilities as $P(\pm 1|a)$, we must have
\begin{eqnarray}
1 & = & P(+1|a) + P(-1|a) \;,\cr
0 & = & P(+1|a) - P(-1|a) \;,
\end{eqnarray}
which yields
\begin{equation}
P(+1|a) \;=\; P(-1|a) \;=\; \dfrac{1}{2}\;.
\end{equation}
Therefore, the measurement of $\GFsigma_1$ on $\ket{a}$ will yield
the two outcomes $+1$ and $-1$ with equal probability, consistent with our earlier results in \cite{Chang:2012eh} and \cite{Chang:2012gg}.
Similarly for the measurement of $\GFsigma_1$ on $\ket{b}$, and
those of $\GFsigma_3$ on $\ket{c}$ or $\ket{d}$.
The expectation values and uncertainties for 
both observables and all states are listed in Table~\ref{V32sigEVs}.

Note that our formalism predicts expectation values but
do not specify the probabilities directly. 
The probabilities must be inferred 
from the expectations values as shown above.
Indeed, though we can write
\begin{eqnarray}
\bra{a}\GFsigma_1\ket{a}
& = & \bra{a}\bigl(\, \1\,\ket{c}\bra{c}-\1\,\ket{d}\bra{d} \,\bigr) \ket{a} \cr
& = & \1\,\braket{a}{c}\braket{c}{a} - \1\,\braket{a}{d}\braket{d}{a} \;,
\end{eqnarray}
we cannot associate $\braket{a}{c}\braket{c}{a}=\braket{a}{d}\braket{d}{a}=-\1$
with the probabilities of the outcomes $\pm 1$.
Furthermore, we will see in the following that
in some cases, the probabilities cannot be uniquely determined from the
expectation values.  
We will argue later that a theory which predicts expectation values
but leaves the probabilities indeterminate still makes perfect physical sense.

\subsection{2D Vector Space over $\bm{GF(9)}$}

\begin{table}[t]
\centering
\begin{tabular}{|c||c|c||c|c||c|c|}
\hline
& $\quad\;\sigma_1\;\;\;$ 
& $\;\;\Delta\sigma_1\;\;$ 
& $\quad\;\sigma_2\;\;$ 
& $\;\;\Delta\sigma_2\;\;\;$ 
& $\quad\;\sigma_3\;\;$ 
& $\;\;\Delta\sigma_3\;\;\;$ 
\\
\hline\hline
$\quad\ket{a}\quad$ 
& $\phantom{-}0$ & $1$ 
& $\phantom{-}0$ & $1$ 
& $\phantom{-}1$ & $0$ 
\\
\hline
$\quad\ket{b}\quad$ 
& $\phantom{-}0$ & $1$ 
& $\phantom{-}0$ & $1$ 
&           $-1$ & $0$ 
\\
\hline
$\quad\ket{c}\quad$ 
& $\phantom{-}1$ & $0$ 
& $\phantom{-}0$ & $1$ 
& $\phantom{-}0$ & $1$ 
\\
\hline
$\quad\ket{d}\quad$ 
&           $-1$ & $0$ 
& $\phantom{-}0$ & $1$ 
& $\phantom{-}0$ & $1$ 
\\
\hline
$\quad\ket{e}\quad$ 
& $\phantom{-}0$ & $1$ 
& $\phantom{-}1$ & $0$ 
& $\phantom{-}0$ & $1$ 
\\
\hline
$\quad\ket{f}\quad$ 
& $\phantom{-}0$ & $1$ 
&           $-1$ & $0$ 
& $\phantom{-}0$ & $1$ 
\\
\hline
\end{tabular}
\caption{Expectation values and uncertainties of spin-like observables
in biorthogonal quantum mechanics on $\V{2}{9}$.}
\label{V92sigEVs}
\end{table}

Next, consider the 2D vector space $\V{2}{9}$
over $GF(9)=\Fthreei$.
This field consists of $3^2=9$ elements given by
$\{\0,\1,-\1,\i,-\i,\1+\i,\1-\i,-\1+\i,-\1-\i\}$.

There are $9^2-1=80$ non-zero vectors in $\V{2}{9}$.
These are the scalar multiples of $80/8=10$ vectors consisting of
the four listed in Eq.~(\ref{abcd}) and the following six:
\begin{equation}
\begin{array}{lll}
\ket{e}=\left[\begin{array}{c} \1 \\ \i \end{array}\right], &
\ket{g}=\left[\begin{array}{c} \1 \\ \1+\i \end{array}\right], &
\ket{i}=\left[\begin{array}{c} \1 \\ -\1+\i \end{array}\right],
\\
\ket{f}=\left[\begin{array}{c} \1 \\ -\i \end{array}\right], &
\ket{h}=\left[\begin{array}{c} \1 \\ \1-\i \end{array}\right], &
\ket{j}=\left[\begin{array}{c} \1 \\ -\1-\i \end{array}\right].
\\
\end{array}
\label{efghkl}
\end{equation}
The dot products of these six vectors with themselves are
\begin{equation}
\begin{array}{lllll}
\dotproduct{\ket{e}}{\ket{e}}& =\;\dotproduct{\ket{f}}{\ket{f}}& =\;-\1\;, & & \\
\dotproduct{\ket{g}}{\ket{g}}& =\;\dotproduct{\ket{h}}{\ket{h}}& =\;
\dotproduct{\ket{i}}{\ket{i}}& =\;\dotproduct{\ket{j}}{\ket{j}}& =\;\0\;.
\end{array}
\end{equation}
As we can see $\ket{g}$, $\ket{h}$, $\ket{i}$, and $\ket{j}$ are all self-orthogonal.
The conjugates of $\ket{e}$ and $\ket{f}$ are
\begin{equation}
\bra{e}\;=\;\bigl[\begin{array}{cc} -\1 &      \i \end{array}\bigr]\;,\qquad
\bra{f}\;=\;\bigl[\begin{array}{cc} -\1 & \!\!-\i \end{array}\bigr]\;.
\end{equation}
Thus, in addition to the two biorthogonal systems listed in Eq.~(\ref{Z3bio}),
$\V{2}{9}^*\times\V{2}{9}$ has a third given by
\begin{equation}
\bigl\{ \{\bra{e},\bra{f}\},\,\{\ket{e},\ket{f}\} \bigr\}\;,
\label{Z9bio}
\end{equation}
and $\ket{e}$ and $\ket{f}$ are added to the list of physical states.

The above biorthogonal system contributes a third operator
to the list of spin-like observables in Eq.~(\ref{Z3obs}):
\begin{equation}
\begin{array}{llll}
\1\;\ket{e}\bra{e} \hspace{-2px}& -\1\;\ket{f}\bra{f} 
& = \;
\left[\begin{array}{rr} \0 & -\i \\ \i & \0 \end{array}\right]
& \equiv \; 
\GFsigma_2
\;.
\end{array}
\end{equation}
By construction,
$\ket{e}$ and $\ket{f}$ are respectively eigenvectors of $\GFsigma_2$ with
eigenvalues $\pm\1$.
The expectation values and uncertainties of all
three observables for all six states are listed in Table~\ref{V92sigEVs}.

\section{Spin Correlations}

In the examples considered above, spin-like observables
were represented by Pauli matrices, with elements in $GP(3^n)$,
acting on the 2D vector spaces $\V{2}{3^n}$, $n=1$ or $2$.
If we associate this model with the spin of one particle,
two particle spin-states will be represented by vectors
in $\V{2}{3^n}\otimes\V{2}{3^n}=\V{4}{3^n}$, $n=1$ or $2$,
while the product spins will be represented by
Kronecker products of the Pauli matrices.
In this section, we will look at the correlations of these
spins.

\subsection{$\bm{n=1}$ case}

The space $\V{4}{3}$ has $3^4-1=80$ non-zero vectors,
every two of which differ by only a multiplicative phase,
namely $-\1$, leaving $80/2=40$ inequivalent vectors.
Of these, $4^2=16$ are products of physical states in $\V{2}{3}$, 
all of which are also physical in $\V{4}{3}$ since
\begin{equation}
(\bra{\psi}\otimes\bra{\phi})(\ket{\psi}\otimes\ket{\phi})
\;=\; \braket{\psi}{\psi}\braket{\phi}{\phi}
\;=\; \1\;,
\end{equation}
if $\braket{\psi}{\psi}=\braket{\phi}{\phi}=\1$.
Of the remaining $40-16=24$ vectors, 
$16$ are self-orthogonal, e.g.
\begin{equation}
\begin{bmatrix}
\;\1\; \\ \;\1\; \\ \;\1\; \\ \;\0\; 
\end{bmatrix} 
\cdot
\begin{bmatrix}
\;\1\; \\ \;\1\; \\ \;\1\; \\ \;\0\; 
\end{bmatrix} 
\;=\; \1+\1+\1+\0
\;=\; \0\;,
\end{equation}
leaving $24-16=8$ physical entangled states.
They are:
\begin{eqnarray}
\ket{S}
& = & \bigl[\begin{array}{cccc} \0 & \1 & -\1 & \0 \end{array}\bigr]^\mathrm{T}\;,\cr
\ket{(ab)}
& = & \bigl[\begin{array}{cccc} \1 & \0 & \0 & -\1 \end{array}\bigr]^\mathrm{T}\;,\cr 
\ket{(cd)}
& = & \bigl[\begin{array}{cccc} \0 & \1 & \1 & \0 \end{array}\bigr]^\mathrm{T}\;,\cr
\ket{(ab)(cd)}
& = & \bigl[\begin{array}{cccc} \1 & \0 & \0 & \1 \end{array}\bigr]^\mathrm{T}\;,\cr 
\ket{(ad)(bc)}
& = & \bigl[\begin{array}{cccc} \1 & \1 & \1 & -\1 \end{array}\bigr]^\mathrm{T}\;,\cr 
\ket{(ac)(bd)}
& = & \bigl[\begin{array}{cccc} -\1 & \1 & \1 & \1 \end{array}\bigr]^\mathrm{T}\;,\cr
\ket{(acbd)}
& = & \bigl[\begin{array}{cccc} \1 & -\1 & \1 & \1 \end{array}\bigr]^\mathrm{T}\;,\cr
\ket{(adbc)}
& = & \bigl[\begin{array}{cccc} \1 & \1 & -\1 & \1 \end{array}\bigr]^\mathrm{T}\;,
\end{eqnarray}
where the labeling is based on the transformation property of each state
under the group of allowed basis transformations $PO(2,3)$. (See Appendix A.1 for details.)

Product spins are represented by $\GFsigma_i\otimes\GFsigma_j$,
$i,j=\mbox{1 or 3}$.
For product states, the expectation value of product spins
factorizes due to the product preserving property of $\ppm$:
\begin{eqnarray}
E(\sigma_i\sigma_j|\psi\phi)
& = & 
\ppm\Bigl[
\bigl(\bra{\psi}\otimes\bra{\phi}\bigr)
\bigl(\GFsigma_i\otimes\GFsigma_j\bigr)
\bigl(\ket{\psi}\otimes\ket{\phi}\bigr)
\Bigr]
\cr
& = &
\ppm\bigl(
\bra{\psi}\GFsigma_i\ket{\psi}
\bra{\phi}\GFsigma_j\ket{\phi}
\bigr)
\cr
& = & 
\ppm\bigl(
\bra{\psi}\GFsigma_i\ket{\psi}
\bigr)
\,
\ppm\bigl(
\bra{\phi}\GFsigma_j\ket{\phi}
\bigr)
\cr
& = &
E(\sigma_i|\psi)\,E(\sigma_j|\phi)
\;.
\end{eqnarray}
This factorization is necessary if we are to
have isolated one particle states.
Again, the product preserving map $\ppm$
plays a fundamental role.
The explicit representations of the product spin
operators are
\begin{eqnarray}
\GFsigma_1\otimes\GFsigma_1
& = &
\left[
\begin{array}{rrrr}
\0 & \0 & \0 & \phantom{-}\1 \\
\0 & \0 & \phantom{-}\1 & \0 \\
\0 & \phantom{-}\1 & \0 & \0 \\
\1 & \0 & \0 & \0
\end{array}
\right],
\cr
\GFsigma_1\otimes\GFsigma_3
& = &
\left[
\begin{array}{rrrr}
\0 &  \0 & \phantom{-}\1 &  \0 \\
\0 &  \0 & \0 & -\1 \\
\1 &  \0 & \0 &  \0 \\
\0 & -\1 & \0 &  \0
\end{array}
\right], 
\cr
\GFsigma_3\otimes\GFsigma_1
& = &  
\left[
\begin{array}{rrrr}
\0 &  \phantom{-}\1 &  \0 &  \0 \\
\1 &  \0 &  \0 &  \0 \\
\0 &  \0 &  \0 & -\1 \\
\0 &  \0 & -\1 &  \0
\end{array}
\right], \cr
\GFsigma_3\otimes\GFsigma_3
& = &  
\left[
\begin{array}{rrrr}
\1 &  \0 &  \0 &  \0 \\
\0 & -\1 &  \0 &  \0 \\
\0 &  \0 & -\1 &  \0 \\
\0 &  \0 &  \0 &  \phantom{-}\1
\end{array}
\right].
\end{eqnarray}
Using these expressions, we can calculate
the spin correlations of this system.

Let us look at what 
the Clauser-Horne-Shimony-Holt (CHSH) bound \cite{Clauser:1969ny} would be.
The CHSH bound is
the upper bound of the absolute value of the following combination of correlators:
\begin{eqnarray}
\lefteqn{C(A,a\,;B,b\,|\Psi)} \cr
& \equiv & E(AB|\Psi)+E(Ab|\Psi)+E(aB|\Psi)-E(ab|\Psi) \;,\qquad
\label{CHSHcorr}
\end{eqnarray}
where $A$ and $a$ are two observables of particle 1, and
$B$ and $b$ are two observables of particle 2.
All four observables are assumed to take on only the values $\pm 1$ upon 
measurement.
For classical hidden variable theory, the bound
on $\left|C(A,a\,;B,b\,|\Psi)\right|$ is 2, while for
canonical QM it is $2\sqrt{2}$ \cite{cirelson}.

In the current case, each of the four observables $A$, $a$, $B$, and $b$
is either $\sigma_1$ or $\sigma_3$.
The cases in which the operators are the negatives of either $\sigma_1$
or $\sigma_3$ need not be considered since
\begin{eqnarray}
\lefteqn{C(A,a\,;B,b\,|\Psi)} \cr
& = & C(A,-a\,;b,B\,|\Psi)
\;=\; -C(-A,a\,;b,B\,|\Psi) \cr
& = & C(a,A\,;B,-b\,|\Psi)
\;=\; -C(a,A\,;-B,b\,|\Psi)\;.\qquad
\end{eqnarray}
To compress our notation, let us define
\begin{equation}
C_{ijk\ell}(\Psi)\;=\;
C(\sigma_i,\sigma_j;\sigma_k,\sigma_\ell|\Psi)\;.
\end{equation}
In the current case, there only four possible combinations
of indices: $C_{1313}$, $C_{1331}$, $C_{3113}$, and $C_{3131}$.
Only the CHSH correlators for entangled states are of interest,
since those for the product states cannot exceed the classical bound.
Furthermore, all eight entangled states can be transformed into the
singlet state $\ket{S}$ by an appropriate local $PO(2,3)$ transformation
so one only needs to consider correlations for this one state.
It is straightforward to show that
\begin{eqnarray}
\bra{S}\GFsigma_1\otimes\GFsigma_1\ket{S}
& = & \bra{S}\GFsigma_3\otimes\GFsigma_3\ket{S}
\;=\; -\1\;,\cr
\bra{S}\GFsigma_1\otimes\GFsigma_3\ket{S}
& = & \bra{S}\GFsigma_3\otimes\GFsigma_1\ket{S}
\;=\; \phantom{-}\0\;.
\label{Scorr}
\end{eqnarray}
From this, we find
\begin{eqnarray}
C_{1313}(S) & = & C_{3131}(S) \;=\;\phantom{-}0\;,\cr
C_{1331}(S) & = & C_{3113}(S) \;=\;-2\;.
\label{CHSH-S}
\end{eqnarray}
Thus, the CHSH bound for this model is the classical 2.

In previous publications \cite{Chang:2012eh,Chang:2012gg}
we argued that the CHSH bound of 2 does not necessarily imply
that the predictions of the model
can be mimicked by a classical hidden variable theory.
In the current case, however, they can be.
Let us denote the classical values of $\sigma_1$ and $\sigma_3$
of particle 1 as $X_1$ and $Z_1$, and those of the 
particle 2 as $X_2$ and $Z_2$, respectively.
The first line of Eq.~(\ref{Scorr}) implies that the
pairs $(X_1,X_2)$ and $(Z_1,Z_2)$ are completely anti-correlated.
Therefore, the only classical configurations possible are 
$(X_1,Z_1;X_2,Z_2)=(+,+;-,-)$,
$(+,-;-,+)$, $(-,+;+,-)$, and $(-,-;+,+)$.
To reproduce the second line of Eq.~(\ref{Scorr}), we only need to 
demand that the probabilities of these configurations satisfy:
\begin{eqnarray}
\dfrac{1}{2}
& = & P(+,+;-,-) + P(-,-;+,+) \cr
& = & P(+,-;-,+) + P(-,+;+,-) \;.
\end{eqnarray}
Thus, an entire class of hidden variable mimics exists.

\subsection{$\bm{n=2}$ case}

The space $\V{4}{9}$ has $9^4-1=6560$ non-zero vectors,
every eight of which differ by only a multiplicative phase, i.e.
an element of $GF(9)\backslash\{0\}$, 
leaving $6560/8=820$ inequivalent states.
Of the $10^2=100$ product states, the
$6^2=36$ products of physical states in $\V{2}{9}$ are also physical in $\V{4}{9}$.
The remaining $64$ product states are self-orthogonal and unphysical.
Of the $820-100=720$ entangled states,
$216$ are self-orthogonal, leaving $720-216=504$ physical
entangled states.
These states fall into three classes that transform among themselves
under local $PU(2,9)$ transformations with 24, 288, and 192 elements each, as explained in Appendix A.2. These classes can be represented by the following three states
%
%
\begin{equation}
\ket{S} = \left[\begin{array}{c} \0 \\ \1 \\ -\1 \\ \0   \end{array}\right],\;\;\;
\ket{T} = \left[\begin{array}{c} \1 \\ \0 \\ \1+\i \\ \1 \end{array}\right],\;\;\;
\ket{U} = \left[\begin{array}{c} \1 \\ \0 \\ \1 \\ \1+\i \end{array}\right],
\end{equation}
with the duals
\begin{eqnarray}
\bra{S} & = & \bigl[\begin{array}{cccc} \0 & -\1 & \1 & \0 \end{array}\bigr]\;,\cr
\bra{T} & = & \bigl[\begin{array}{cccc} \1 & \0 & \1-\i & \1 \end{array}\bigr]\;,\cr
\bra{U} & = & \bigl[\begin{array}{cccc} \1 & \0 & \1 & \1-\i \end{array}\bigr]\;.
\end{eqnarray}
Thus, we only need to calculate the correlators for these states to obtain the CHSH bound.
Since there are three spin observables $\GFsigma_1$, $\GFsigma_2$, and $\GFsigma_3$
this time, the number of possible CHSH correlators is $6^2=36$.


Let us first look at the correlators involving only $\GFsigma_1$ and $\GFsigma_3$.
The correlations for the state $\ket{S}$ are the same as those listed in 
Eq.~(\ref{Scorr}) and (\ref{CHSH-S}).
Those for the state $\ket{T}$ are
\begin{eqnarray}
\bra{T}\GFsigma_1\otimes\GFsigma_1\ket{T}
& = & \bra{T}\GFsigma_1\otimes\GFsigma_3\ket{T}
\;=\; -\1\;,\cr
\bra{T}\GFsigma_3\otimes\GFsigma_1\ket{T}
& = & \1\;,\cr
\bra{T}\GFsigma_3\otimes\GFsigma_3\ket{T}
& = & \0\;,
\label{Tcorr}
\end{eqnarray}
from which we obtain
\begin{eqnarray}
C_{1313}(T) & = & -1\;,\cr
C_{3113}(T) & = & C_{3131}(T) \;=\; 1\;,\cr
C_{1331}(T) & = & -3\;.
\end{eqnarray}
Similarly, for the state $\ket{U}$ we have
\begin{eqnarray}
\bra{U}\GFsigma_1\otimes\GFsigma_1\ket{U}
& = & \bra{U}\GFsigma_1\otimes\GFsigma_3\ket{U} \cr
& = & \bra{U}\GFsigma_3\otimes\GFsigma_3\ket{U}
\;=\; -\1\;,\cr
\bra{U}\GFsigma_3\otimes\GFsigma_1\ket{U}
& = & \1\;,
\label{Ucorr}
\end{eqnarray}
and
\begin{eqnarray}
C_{1313}(U) & = & 
C_{3113}(U) \;=\; C_{3131}(U) \;=\; 0\;,\cr
C_{1331}(U) & = & -4\;.
\end{eqnarray}
As can be seen, the absolute value of the correlator $C_{1331}$
for the states $\ket{T}$ and $\ket{U}$ exceed not only the 
classical bound of 2 but also the Cirel'son bound of $2\sqrt{2}$.
In a similar fashion, we have scanned 
all 36 spin combinations for the three states and
have obtained the tally shown in Table~\ref{STUcorrTally}.
Thus, we find that the CHSH bound for this model is 4.

\begin{table}
\centering
\begin{tabular}{|c||c|c|c|c|c|}
\hline
\ state\ \ & $\quad 0\quad$ & $\quad 1\quad$ & $\quad 2\quad$ & $\quad 3\quad$ & $\quad 4\quad$ \\
\hline\hline
$\ket{S}$ & 6 & 24 & 6 & 0 & 0 \\
\hline
$\ket{T}$ & 6 & 18 & 6 & 6 & 0 \\
\hline
$\ket{U}$ & 12 & 12 & 4 & 4 & 4 \\
\hline
\end{tabular}
\caption{The number of CHSH correlators with the respective absolute values for the
three states $\ket{S}$, $\ket{T}$, and $\ket{U}$.}
\label{STUcorrTally}
\end{table}

Unlike the $n=1$ case, which had a CHSH bound of 2,
the above correlations cannot be reproduced by
any classical hidden variable theory.  
For instance, the first line of Eq.~(\ref{Tcorr}) demands that the pairs
$(X_1,X_2)$ and $(X_1,Z_2)$ are completely anti-correlated,
while the second line demands that the pair $(Z_1,X_2)$ is completely correlated.
But then $X_1=\pm 1$ would imply $X_2=\mp 1$ and $Z_2=\mp 1$, 
the first of which implies $Z_1=\mp 1$. Therefore, the pair
$(Z_1,Z_2)$ must also be completely correlated which contradicts
the third line of Eq.~(\ref{Tcorr}).
Similarly, Eq.~(\ref{Ucorr}) demands that the pairs
$(X_1,X_2)$, $(X_1,Z_2)$, and $(Z_1,Z_2)$ are completely
anti-correlated, while $(Z_1,X_2)$ is completely correlated.
But then $X_1=\pm 1$ would imply $X_2=\mp 1$ and $Z_2=\mp 1$, 
the former of which implies $Z_1=\mp 1$ while the
latter $Z_1=\pm 1$, leading to a contradiction.
Of course, this is not surprising since the CHSH bound for
classical hidden variable theories is 2.
The unexpected result is that the CHSH bound of our model also
exceeds the quantum Cirel'son bound of $2\sqrt{2}$.
In the next section, we will take a careful look 
at how this comes about.

\section{Expectation Values without Definite Probabilities}

In canonical QM, the states that correspond to 
$\ket{S}$, $\ket{T}$, and $\ket{U}$ are
%
%
\begin{equation}
\ket{\tilde{S}} = \dfrac{1}{\sqrt{2}}\!
\left[\begin{array}{c} 0 \\ 1 \\ -1 \\ 0   \end{array}\right]\!,\;\,
\ket{\tilde{T}} = \dfrac{1}{2}\!
\left[\begin{array}{c} 1 \\ 0 \\ 1+i \\ 1 \end{array}\right]\!,\;\,
\ket{\tilde{U}} = \dfrac{1}{2}\!
\left[\begin{array}{c} 1 \\ 0 \\ 1 \\ 1+i \end{array}\right]\!,
\end{equation}
Calculating the correlations of canonical spin $\Csigma_i$
for the state $\ket{\tilde{S}}$ in canonical QM, we find
\begin{eqnarray}
\bra{\tilde{S}}\Csigma_1\otimes\Csigma_1\ket{\tilde{S}}
& = & \bra{\tilde{S}}\Csigma_3\otimes\Csigma_3\ket{\tilde{S}}
\;=\; -1\;,\cr
\bra{\tilde{S}}\Csigma_1\otimes\Csigma_3\ket{\tilde{S}}
& = & \bra{\tilde{S}}\Csigma_3\otimes\Csigma_1\ket{\tilde{S}}
\;=\; \phantom{-}0\;,
\label{ScorrCQM}
\end{eqnarray}
which agree with those for $\ket{S}$ in Eq.~(\ref{Scorr}) via the 
product preserving map $\ppm$.
For $\ket{\tilde{T}}$ and $\ket{\tilde{U}}$, however, we find:
\begin{eqnarray}
\bra{\tilde{T}}\Csigma_1\otimes\Csigma_1\ket{\tilde{T}}
& = & \bra{\tilde{T}}\Csigma_1\otimes\Csigma_3\ket{\tilde{T}}
\;=\; \dfrac{1}{2}\;,\cr
\bra{\tilde{T}}\Csigma_3\otimes\Csigma_1\ket{\tilde{T}}
& = & -\dfrac{1}{2}\;,\cr
\bra{\tilde{T}}\Csigma_3\otimes\Csigma_3\ket{\tilde{T}}
& = & 0\;,
\cr
& & \cr
\bra{\tilde{U}}\Csigma_1\otimes\Csigma_1\ket{\tilde{U}}
& = & \bra{\tilde{U}}\Csigma_1\otimes\Csigma_3\ket{\tilde{U}} \cr
& = & \bra{\tilde{U}}\Csigma_3\otimes\Csigma_3\ket{\tilde{U}}
\;=\; \dfrac{1}{2}\;,\cr
\bra{\tilde{U}}\Csigma_3\otimes\Csigma_1\ket{\tilde{U}}
& = & -\dfrac{1}{2}\;,
\end{eqnarray}
Thus, the correspondence here is
\begin{equation}
-\1 \;\leftrightarrow\; \dfrac{1}{2}\;,\qquad
\1 \;\leftrightarrow\; -\dfrac{1}{2}\;,
\end{equation}
which is to be expected since $\1\div\2=\2=-\1$ in $GF(3)$.
So the large correlation is due to the fact that
$GF(3)$ has only three elements $\{-\1,\0,\1\}$ which are
mapped to $\{-1,0,1\}\in\R$ by the product preserving map $\ppm$.
The fact that the only spin-correlations possible are $0$ or $\pm 1$
will of course persist for larger values of $p=3\mod 4$ as long
as we use $\ppm$.

\begin{table}[b]
\centering
\begin{tabular}{|c||c|c|c|c||c|}
\hline
& $\quad ++\quad$ & $\quad +-\quad$ & $\quad -+\quad$ & $\quad --\quad$ & \ \ E.V.\ \ \ \\
\hline\hline
$\phantom{\bigg|}\quad\ket{\tilde{S}}\quad$ & $0$ & $\dfrac{1}{2}$ & $\dfrac{1}{2}$ & $0$ & $-1$\\
\hline
$\phantom{\bigg|}\ket{\tilde{T}}$ & $\dfrac{1}{4}$ & $0$ & $\dfrac{1}{2}$ & $\dfrac{1}{4}$ & $0$ \\
\hline
$\phantom{\bigg|}\ket{\tilde{U}}$ & $\dfrac{1}{4}$ & $0$ & $\dfrac{1}{4}$ & $\dfrac{1}{2}$ & $+\dfrac{1}{2}$ \\
\hline
\end{tabular}
\caption{The probabilities of the four possible outcomes $++$, $+-$, $-+$, and $--$
in canonical quantum mechanic when $\Csigma_3\otimes\Csigma_3$ is measured on 
the canonical states $\ket{\tilde{S}}$, $\ket{\tilde{T}}$, and $\ket{\tilde{U}}$.
}
\label{QMProbs}
\end{table}

What are the corresponding probabilities?  Let us take
the spins in the $Z$-direction,
$\sigma_3\otimes\sigma_3$, as an example.  
The probabilities of the
outcomes $(\sigma_3\sigma_3)=(++)$, $(+-)$, $(-+)$, and $(--)$ 
in canonical QM are listed in Table~\ref{QMProbs}.
As can be seen, they reproduce the correlations listed above
as they should.

In our `mutant' biorthogonal quantum mechanics, however,
the probabilities of individual outcomes are ill defined as discussed above. 
Taking the point of view that the probabilities must be inferred from 
the expectation values, we have the constraints
\begin{equation}
\begin{array}{l}
P(++|T)+P(+-|T)+P(-+|T)+P(--|T) = 1\;,\\
P(++|T)-P(+-|T)-P(-+|T)+P(--|T) = 0\;,
\end{array}
\end{equation}
for $\ket{T}$, and
\begin{equation}
\begin{array}{l}
P(++|U)+P(+-|U)+P(-+|U)+P(--|U) = 1\;,\\
P(++|U)-P(+-|U)-P(-+|U)+P(--|U) = -1\;,
\end{array} 
\end{equation}
for $\ket{U}$.
These constraints imply
\begin{eqnarray}
\dfrac{1}{2}
& = & P(++|T)+P(--|T) \cr
& = & P(+-|T)+P(-+|T) \;,\phantom{\bigg|}\cr
0 & = & P(++|U)+P(--|U) \;,\phantom{\bigg|}\cr
1 & = & P(+-|U)+P(-+|U) \;,
\end{eqnarray}
but beyond this the probabilities cannot be specified. 
Therefore, though our formalism predicts definite
expectation values, it leaves probabilities indeterminate.
Physically, we interpret this to mean that if the same measurement
is repeated many times, the average of the outcomes will 
converge to the predicted expectation value, while the frequencies of
each outcome will continue to fluctuate.

This indeterminacy is characteristic of the approach used here, and can be understood more generally by re-examining the defining relation between expectation values and probability distributions.  In conventional QM, it is possible to construct the probability
distribution for the measurement outcomes of some observable through 
use of the system of equations formed by the expectation values of the powers 
of the observable in question. This is not possible for spin observables in the model 
under consideration due to the cyclic nature of the underlying field. Explicitly, the system of equations:
 \begin{eqnarray}
E(A|\psi)   & = & \sum_{\alpha}\alpha P(\alpha|\psi) \;,\cr
E(A^2|\psi) & = & \sum_{\alpha}\alpha^2 P(\alpha|\psi) \;, \cr
& \vdots & \cr
E(A^N|\psi) & = & \sum_{\alpha}\alpha^N P(\alpha|\psi) \;,
\end{eqnarray}
will be singular if $N$ is greater than the least common multiple of the multiplicative orders of 
the eigenvalues $\alpha$ of $\hat{A}$ since the cyclic nature of the field is necessarily shared by the 
eigenvalues  when the product preserving map also preserves the eigenvalues. In our examples,
using $GF(3)$ as the `real' field, the eigenvalues of spin observables, $\{+1,-1\}$, have multiplicative orders no 
greater than 2. Thus, when we form a four level system by entangling two particles, we find that the system 
of equations needed to solve for the probabilities of these four measurement outcomes is singular and 
cannot be used to assign consistent probabilities.

In Ref.~\cite{Chang:2011yt}, we conjectured that a `doubly' quantized
theory may predict super-quantum correlations with a CHSH bound which
exceeds the Cirel'son value of $2\sqrt{2}$.  
A state in such a theory can be thought of as a `superposition' of various
`singly' quantized states, each of which predicts definite probabilities.
A `measurement' in a `doubly' quantized theory can be expected to
collapse the `doubly' quantized state to a `singly' quantized one, selecting
a particular probability distribution from all possible ones.
Every `measurement' will lead to a different probability distribution,
so no definite probability will be predicted.
These considerations suggest that biorthogonal QM
is a candidate model for such a `doubly' quantized theory.

\section{Discussion}

One of the simplest realization of how quantum theory differs from
its classical counterpart is given by the celebrated Bell inequalities,
or its slightly generalized version, the CHSH inequalities
\cite{Clauser:1969ny,bell,GHZ,Hardy:1993zza,cirelson}. According to these
inequalities the classical
and quantum physics are clearly separated by $O(1)$ effects. It has been pointed out in the literature that the purely statistical reasoning
leads to the maximal ``super-quantum bound'' of 4 \cite{super}. In one of our previous papers
we have pointed out the special nature of such a super-quantum theory \cite{Chang:2011yt}.
Given the fact that the CHSH inequalities rely on the knowledge of
expectation values (and not probabilities) in this paper we have focused  on the
requirement that expectation values of a super-quantum theory should
satisfy the bound of 4.

Note that our present work is distinguished from other efforts that try
to eliminate theories which violate the quantum bound or which claim the uniqueness
of the canonical complex quantum theory because of the supposed
unphysical nature of super-quantum theories (see \cite{trivial}).
As is well known, the expectation values and the probabilities are
related by a quadratic map in canonical quantum theories and its
real counterparts \cite{Stueckelberg:1960}.
That this map is quadratic can be argued on general grounds,
and the robustness of the Born rule \cite{triple}, by pointing
out the generic nature of the Fisher metric on the space of measured events \cite{wootters}.

The CHSH observable relies only on the computation of the expectation values.
In order to achieve the super-quantum bound of 4, one immediately realizes (at least on a heuristic level)  that the expectation values should be ``mutated'' so that the last term in the CHSH observable changes its sign. Given the canonical relation between the expectation values and the probabilities, such a ``mutation'' of the computation of the expectation values would, at least naively, influence the probabilities as well. This is precisely what we find in a concrete mathematical model explored in this paper: the CHSH observable computed in the mutant quantum mechanics over the finite field $GF(9)$ is explicitly equal to 4, which in turn implies that the probabilities are indeterminate in such a super-quantum theory.

Indeterminate probabilities are a consequence of our construction and, in this particular case, a necessary feature of such a super-quantum theory.
Note that this statement also goes against some efforts in the foundations
of quantum theory, which try to base the canonical complex quantum theory
solely on the concept of probability (see for example \cite{fuchs}). Our point is that even though canonical quantum theory might
be solely based on the concept of probability, super-quantum theory
does not have to be. This reinforces the experience of modern QFT
(especially the conformal QFT's) in which one operates only with correlation functions.

In Appendix A, we show that, in the context of Galois biorthogonal QM,
the projective orthogonal and the projective
unitary groups play the natural role of the orthogonal and unitary groups of canonical QM.
This maintains a parallel with our previous papers 
on Galois field QM \cite{Chang:2012eh,Chang:2012gg}
where we have shown that the complex (and real) projective spaces, 
which define the geometry of canonical quantum theory,
can be naturally replaced by their finite projective counterparts.
Similarly, in this work, the orthogonal and unitary groups that define
the invariance of expectation values in the real and complex quantum theories
are replaced by their projective counterparts.
It is of course tempting to contemplate that the
general structure of biorthogonal systems, the graded valuation of expectation
values, and the indeterminate nature of probabilities is valid for more
general constructions of super-quantum theories, including the ones that we
expect to be relevant in quantum theory of gravity.

To summarize: in this paper we have presented perhaps the simplest model for
quantum super-correlations. Quantum super-correlations are realized in
the model together with a signature feature:
the physics of the model is entirely determined in terms of expectation values,
whereas the probabilities are, in general, indeterminate.
This feature is actually quite
natural (and desirable) from various point of view suggested by different modern avenues of fundamental physics.

We note that the fact that the probabilities are indeterminate in our explicit construction
also meshes well with some expectations from various attempts at quantum theory of gravity (including the ones in which conformal field theories
are used to define a quantum theory of gravity in particular asymptotic geometries.)
Indeed, that fundamental quantum theories can be defined in terms of expectation values (which is most obvious in the path integral formulation), is a feature found in modern conformal field theories, which are quantum field theory formulated from a purely algebraic viewpoint, without the use of Lagrangians (or Hamiltonians) or Feynman rules. For example, the familiar S-matrix of the
canonical quantum field theory, which comes about from compounding expectation values
(correlation functions) with wave-functions of external probes, is not a well-defined concept in conformal field theory. As is well known, conformal field theories, can be dual to (quantum) gravitational theories in certain
background (the AdS spaces \cite{adscft}, and also in the context of the
observed cosmological de Sitter spacetimes \cite{dscft}). 

Thus, this feature should be relevant in the context of quantum gravity as well.
Indeed, different approaches to non-perturbative quantum gravity and quantum cosmology \cite{gh,Hardy:2005fq,chia}, suggest that the individual probability for specific measurements could be
indeterminate, and that the observables in that context are different from
the usual observables found in the canonical quantum theory.
The model considered here should be viewed as a concrete realization of this
general expectation.

The model sheds new light on the foundations of quantum theory, 
and attempts to understand the simplest set of reasonable axioms that
lead to canonical quantum theory, which could lead to natural generalizations of quantum theory expected in the context of quantum theory of gravity \cite{chia,Hardy:2001jk}.

Finally, we note that this work presents an alternative pathway to 
constructing a quantum theory on a vector space without an inner product 
from the one introduced in Refs.~\cite{Chang:2012eh,Chang:2012gg}. 
Application of the two constructions to Banach spaces \cite{banach}
would be a natural place to further clarify the difference between the
two approaches, do away with the product preserving map from $GF(p)$ to $\R$,
and search for models which may serve as closer representations of reality
where various quantum gravitational ideas discussed above can be explored.

We will return to these, and related issues in future works.

\begin{acknowledgement}
We would like to thank Rafael Sorkin and Chia Tze for informative discussions.  
ZL, DM and TT are supported in part by
the U.S. Department of Energy, grant DE-FG05-92ER40677, task A.
DM thanks the Perimeter Institute and the Aspen Center for Physics for providing stimulating
working environments during the completion of this paper.
\end{acknowledgement}

\appendix
\section{Group of Basis Transformations}

\subsection{$\V{2}{3}$ case}

There are only two biorthogonal systems
in $\V{2}{3}^*\times\V{2}{3}$ listed in Eq.~(\ref{Z3bio}),
up to ordering of the vectors and multiplicative phases.
Thus, the allowed bases of $\V{2}{3}$ are 
\begin{equation}
\begin{array}{ll}
\pm\,\{\,\ket{a},\pm\ket{b}\,\}\,,&\quad 
\pm\,\{\,\ket{b},\pm\ket{a}\,\}\,,\\
\pm\,\{\,\ket{c},\pm\ket{d}\,\}\,,&\quad
\pm\,\{\,\ket{d},\pm\ket{c}\,\}\,.
\end{array}
\label{V23bases}
\end{equation}
Thus, the group of all possible basis transformations consist of 
sixteen matrices given by
\begin{equation}
\begin{array}{rlrl}
e&\leftrightarrow\;
\pm\left[\begin{array}{rr}  \1 & \phantom{-}\0 \\  \0 &  \1 \end{array}\right],\quad &
(ab)&\leftrightarrow\;
\pm\left[\begin{array}{rr}  \0 & \phantom{-}\1 \\ \1 &  \0 \end{array}\right],\\
(cd)&\leftrightarrow\;
\pm\left[\begin{array}{rr}  \1 & \phantom{-}\0 \\ \0 & -\1 \end{array}\right],\quad &
(ab)(cd)&\leftrightarrow\;
\pm\left[\begin{array}{rr}  \0 & -\1 \\ \1 &  \0 \end{array}\right],\\
(ac)(bd)&\leftrightarrow\;
\pm\left[\begin{array}{rr}  \1 &  \1 \\ \1 & -\1 \end{array}\right],\quad &
(ad)(bc)&\leftrightarrow\;
\pm\left[\begin{array}{rr} -\1 & \phantom{-}\1 \\  \1 &  \1 \end{array}\right],\\
(acbd)&\leftrightarrow\;
\pm\left[\begin{array}{rr}  \1 & -\1 \\  \1 &  \1 \end{array}\right],\quad &
(adbc)&\leftrightarrow
\;\pm\left[\begin{array}{rr}  \1 & \phantom{-}\1 \\ -\1 &  \1 \end{array}\right].
\end{array}
\label{PO23D4}
\end{equation}
However, since we identify vectors that only differ by multiplicative phases as
representing the same physical state, we identify the matrices that only differ
by a multiplicative phase as representing the same transformation on
the projective space $PG(1,3)$,
each of which corresponds to a permutation of the vector labels $a$, $b$, $c$, and $d$
as indicated above.
These eight transformations constitute the projective orthogonal group $PO(2,3)\cong D_4$,
namely, the group of $2\times 2$ matrices $\underline{O}$ with elements in $GF(3)$ which satisfy
the condition
\begin{equation}
\underline{O}^\mathrm{T}\underline{O}\;=\;\pm\mathbf{\1}_{2\times 2}\;,
\end{equation}
with matrices which differ by a sign identified.
This group is a subgroup of the projective general linear group $PGL(2,3)\cong S_4$.

\begin{figure}[t]
\sidecaption
\includegraphics[width=6cm]{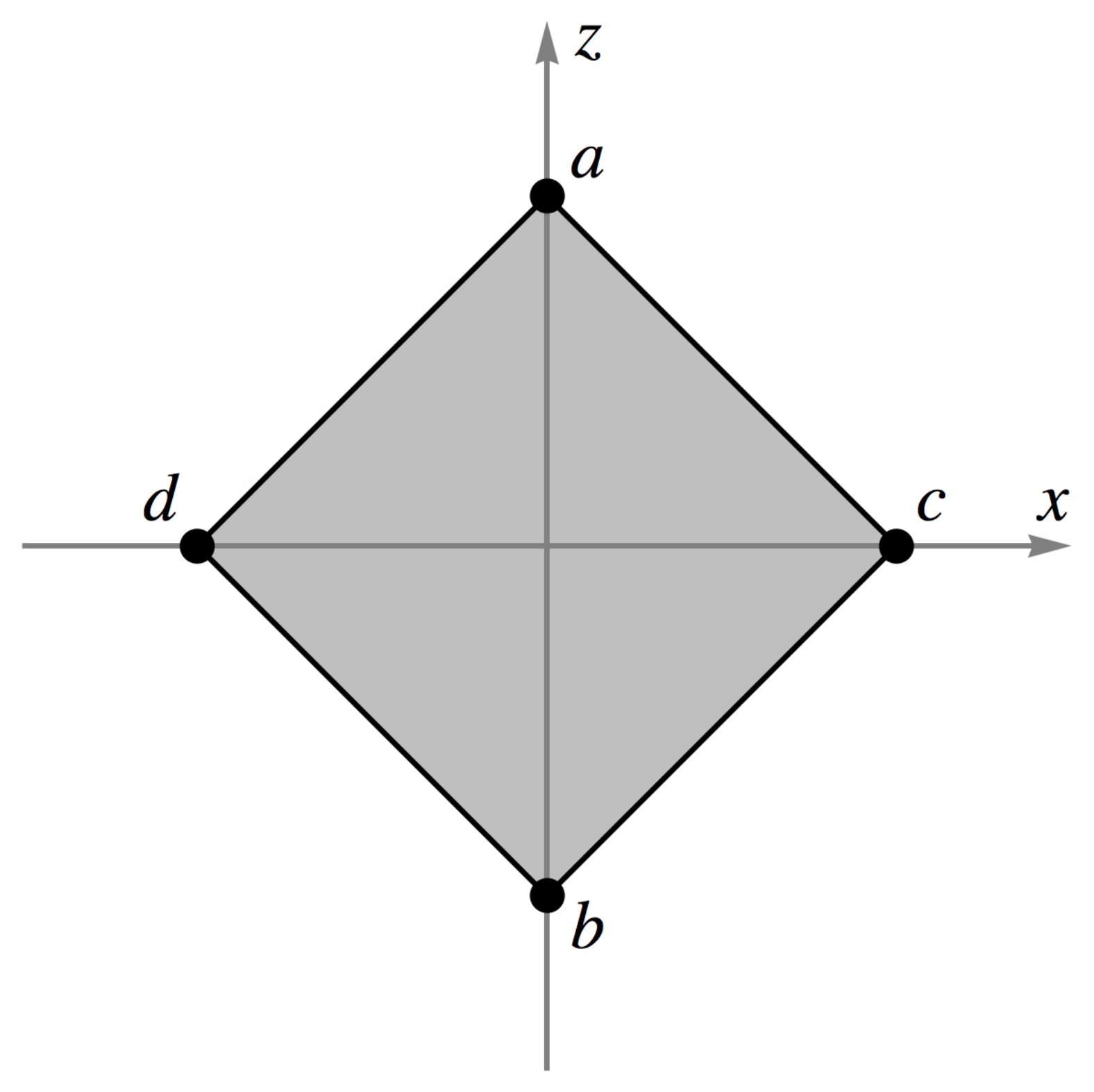}
\caption{The correspondence between the dihedral group $D_4$ 
and the projective orthogonal group $PO(2,3)$.
Every $D_4$ rotation of the quadrangle corresponds to a
permutation of the four vertex labels $abcd$ belonging to $PO(2,3)$.
}
\label{D4fig}
\end{figure}

The isomorphism between $PO(2,3)$ and $D_4$ is implemented by labeling the four corners of a
square as shown in Fig.~\ref{D4fig}.
Every rotation of the quadrangle in $D_4$ leads to a permutation
of the four vertex labels, which is the corresponding element of $PO(2,3)$.
The two spin observables $\GFsigma_1$ and $\GFsigma_3$ transform under
$PO(2,3)$ permutations as
\begin{eqnarray}
e\; & : & 
\GFsigma_1\rightarrow \GFsigma_1\;,\;
\GFsigma_3\rightarrow \GFsigma_3\;,
\cr
(ab) & : & 
\GFsigma_1\rightarrow \GFsigma_1\;,\;
\GFsigma_3\rightarrow -\GFsigma_3\;,
\cr
(cd) & : & 
\GFsigma_1\rightarrow -\GFsigma_1\;,\;
\GFsigma_3\rightarrow \GFsigma_3\;,
\cr
(ab)(cd) & : & 
\GFsigma_1\rightarrow -\GFsigma_1\;,\;
\GFsigma_3\rightarrow -\GFsigma_3\;,
\cr
(ac)(bd) & : & 
\GFsigma_1\rightarrow \GFsigma_3\;,\;
\GFsigma_3\rightarrow \GFsigma_1\;,
\cr
(ad)(bc) & : & 
\GFsigma_1\rightarrow -\GFsigma_3\;,\;
\GFsigma_3\rightarrow -\GFsigma_1\;,
\cr
(acbd) & : & 
\GFsigma_1\rightarrow -\GFsigma_3\;,\;
\GFsigma_3\rightarrow \GFsigma_1\;,
\cr
(adbc) & : & 
\GFsigma_1\rightarrow \GFsigma_3\;,\;
\GFsigma_3\rightarrow -\GFsigma_1\;,
\end{eqnarray}
just as they should under rotations of the quadrangle.

The eight elements of $PO(2,3)$ fall into five conjugacy classes given by
\begin{eqnarray}
& & \{e\}\;,\;\{(ab)(cd)\}\;,\;\{(ab),(cd)\}\;,\cr 
& & \{(ac)(bd),(ad)(bc)\}\;,\; \mbox{and}\; \{(acbd),(adbc)\}\;.
\label{D4classes}
\end{eqnarray}
The eight physical entangled states in $\V{2}{3}\times\V{2}{3}=\V{4}{3}$ 
also fall into five classes that
transform among themselves under global $PO(2,3)$. 
They can be classified and labeled according to their transformation properties 
under the full global $PGL(2,3)$.
\begin{eqnarray}
\ket{S}
& = & \bigl[\begin{array}{cccc} \0 & \1 & -\1 & \0 \end{array}\bigr]^\mathrm{T}\;,\cr
\ket{(ab)}
& = & \bigl[\begin{array}{cccc} \1 & \0 & \0 & -\1 \end{array}\bigr]^\mathrm{T}\;,\cr 
\ket{(cd)}
& = & \bigl[\begin{array}{cccc} \0 & \1 & \1 & \0 \end{array}\bigr]^\mathrm{T}\;,\cr
\ket{(ab)(cd)}
& = & \bigl[\begin{array}{cccc} \1 & \0 & \0 & \1 \end{array}\bigr]^\mathrm{T}\;,\cr 
\ket{(ad)(bc)}
& = & \bigl[\begin{array}{cccc} \1 & \1 & \1 & -\1 \end{array}\bigr]^\mathrm{T}\;,\cr 
\ket{(ac)(bd)}
& = & \bigl[\begin{array}{cccc} -\1 & \1 & \1 & \1 \end{array}\bigr]^\mathrm{T}\;,\cr
\ket{(acbd)}
& = & \bigl[\begin{array}{cccc} \1 & -\1 & \1 & \1 \end{array}\bigr]^\mathrm{T}\;,\cr
\ket{(adbc)}
& = & \bigl[\begin{array}{cccc} \1 & \1 & -\1 & \1 \end{array}\bigr]^\mathrm{T}\;. 
\end{eqnarray}
Here, $\ket{S}$ is the singlet state which is invariant under all transformations
in $PGL(2,3)$. The state $\ket{(ab)(cd)}$ is also a singlet under $PO(2,3)$ 
transformations, but transforms into $\ket{(ac)(bd)}$ and $\ket{(ad)(cd)}$
under the full $PGL(2,3)$.
The other states transform in pairs under $PO(2,3)$, 
falling into the same classes as the $PO(2,3)$ transformations themselves 
as listed in Eq.~(\ref{D4classes}).

Under local $PO(2,3)$ transformations, that is, $PO(2,3)$ transformations acting
on only one of the $\V{2}{3}$ vector spaces in $\V{4}{3}=\V{2}{3}\times\V{2}{3}$, 
all eight states fall into the same class and can be transformed into the
singlet state $\ket{S}$.  Explicitly, we have:
\begin{eqnarray}
\ket{S}
& = & (cd)_1\ket{(cd)} \;=\; (cd)_2\ket{(cd)}\cr
& = & (ab)_1\ket{(ab)} \;=\; (ab)_2\ket{(ab)} \cr
& = & (ab)_1(cd)_1\ket{(ab)(cd)} \;=\; (ab)_2(cd)_2\ket{(ab)(cd)}\cr
& = & (ac)_1(bd)_1\ket{(ac)(bd)} \;=\; (ac)_2(bd)_2\ket{(ac)(bd)} \cr
& = & (ad)_1(bc)_1\ket{(ad)(bc)} \;=\; (ad)_2(bc)_2\ket{(ad)(bd)} \cr
& = & (acbd)_1\ket{(acbd)} \;=\; (acbd)_2\ket{(adbc)} \cr
& = & (adbc)_1\ket{(adbc)} \;=\; (adbc)_2\ket{(acbd)} \;, 
\end{eqnarray}
where the subscript indicates which $\V{2}{3}$ space the transformations are
acting on.

The above considerations indicate that it suffices to calculate
the CHSH bound for only the singlet state $\ket{S}$.

\subsection{$\V{2}{9}$ case}

In $\V{2}{9}$, we have three biorthogonal systems listed in
Eqs.~(\ref{Z3bio}) and (\ref{Z9bio}).
Unlike $\V{2}{3}$, this space has unphysical self-orthogonal
vectors so some care is necessary in listing possible bases
since the dot product is not invariant under generic basis transformations,
and a non-self-orthogonal vector may be mapped to a self-orthogonal one.
Using the notation of Eqs.~(\ref{abcd}) and (\ref{efghkl}),
the allowed bases are
\begin{equation}
\begin{array}{ll}
\eta\,\{\,\ket{a},\pm\ket{b}\,\}\,,&\quad 
\eta\,\{\,\ket{a},\pm\i\ket{b}\,\}\,,\\
\eta\,\{\,\ket{b},\pm\ket{a}\,\}\,,&\quad 
\eta\,\{\,\ket{b},\pm\i\ket{a}\,\}\,,\\
\eta\,\{\,\ket{c},\pm\ket{d}\,\}\,,&\quad 
\eta\,\{\,\ket{c},\pm\i\ket{d}\,\}\,,\\
\eta\,\{\,\ket{d},\pm\ket{c}\,\}\,,&\quad 
\eta\,\{\,\ket{d},\pm\i\ket{c}\,\}\,,\\
\eta\,\{\,\ket{e},\pm\ket{f}\,\}\,,&\quad 
\eta\,\{\,\ket{e},\pm\i\ket{f}\,\}\,,\\
\eta\,\{\,\ket{f},\pm\ket{e}\,\}\,,&\quad 
\eta\,\{\,\ket{f},\pm\i\ket{e}\,\}\,,
\end{array}
\label{V29bases}
\end{equation}
where $\eta$ is an arbitrary phase, that is,
an element of $GF(9)\backslash\{0\}$.
Thus, the group of allowed basis
transformation are represented by the following matrices:
\begin{equation}
\begin{array}{rlrl}
e&\!\leftrightarrow
\eta\left[\begin{array}{rr}  \1 & \phantom{-}\0 \\  \0 &  \1 \end{array}\right], &
(ab)(ef)&\!\leftrightarrow
\eta\left[\begin{array}{rr}  \0 & \phantom{-}\1 \\ \1 &  \0 \end{array}\right],\\
(cd)(ef)&\!\leftrightarrow
\eta\left[\begin{array}{rr}  \1 & \phantom{-}\0 \\ \0 & -\1 \end{array}\right], &
(ab)(cd)&\!\leftrightarrow
\eta\left[\begin{array}{rr}  \0 & -\1 \\ \1 &  \0 \end{array}\right],
\\
(acbd)&\!\leftrightarrow
\eta\left[\begin{array}{rr}  \1 & -\1 \\  \1 &  \1 \end{array}\right], &\qquad
(ac)(bd)(ef)&\!\leftrightarrow
\eta\left[\begin{array}{rr}  \1 &  \1 \\ \1 & -\1 \end{array}\right], \\
(adbc)&\!\leftrightarrow
\eta\left[\begin{array}{rr}  \1 & \phantom{-}\1 \\ -\1 &  \1 \end{array}\right], &
(ad)(bc)(ef)&\!\leftrightarrow
\eta\left[\begin{array}{rr} -\1 & \phantom{-}\1 \\  \1 &  \1 \end{array}\right],\\
(aebf)&\!\leftrightarrow
\eta\left[\begin{array}{rr}  \1 & \phantom{-}\i \\ \i &  \1 \end{array}\right], &
(ae)(bf)(cd)&\!\leftrightarrow\;
\eta\left[\begin{array}{rr}  \1 & -\i \\ \i & -\1 \end{array}\right],\\
(afbe)&\!\leftrightarrow
\eta\left[\begin{array}{rr}  \1 & -\i \\ -\i & \1 \end{array}\right], &
(af)(be)(cd)&\!\leftrightarrow\;
\eta\left[\begin{array}{rr}  \1 & \i \\ -\i & -\1 \end{array}\right],\\
(cedf)&\!\leftrightarrow
\eta\left[\begin{array}{rr}  \1 & \phantom{-}\0 \\  \0 & \i \end{array}\right], &
(ab)(ce)(df)&\!\leftrightarrow\;
\eta\left[\begin{array}{rr}  \0 & -\i \\ \1 &  \0 \end{array}\right],\\
(cfde)&\!\leftrightarrow
\eta\left[\begin{array}{rr}  \1 & \phantom{-}\0 \\ \0 & -\i \end{array}\right], &
(ab)(cf)(de)&\!\leftrightarrow\;
\eta\left[\begin{array}{rr}  \0 & \phantom{-}\i \\ \1 &  \0 \end{array}\right],
\\
(ace)(bdf)&\!\leftrightarrow
\eta\left[\begin{array}{rr}  \1 & -\i \\  \1 &  \i \end{array}\right], &
(adf)(bce)&\!\leftrightarrow
\eta\left[\begin{array}{rr}  \1 & \phantom{-}\i \\ -\1 & \i \end{array}\right],\\
(acf)(bde)&\!\leftrightarrow
\eta\left[\begin{array}{rr}  \1 &  \i \\ \1 & -\i \end{array}\right], &
(ade)(bcf)&\!\leftrightarrow
\eta\left[\begin{array}{rr} -\1 & \phantom{-}\i \\ \1 &  \i \end{array}\right],\\
(aec)(bfd)&\!\leftrightarrow
\eta\left[\begin{array}{rr}  \1 &  \1 \\ \i & -\i \end{array}\right], &
(afd)(bec)&\!\leftrightarrow
\eta\left[\begin{array}{rr} -\1 & \phantom{-}\1 \\  \i &  \i \end{array}\right],\\
(aed)(bfc)&\!\leftrightarrow
\eta\left[\begin{array}{rr}  \1 & -\1 \\  \i &  \i \end{array}\right], &
(afc)(bed)&\!\leftrightarrow
\eta\left[\begin{array}{rr}  \1 & \phantom{-}\1 \\ -\i &  \i \end{array}\right],\\
\end{array}
\label{PO29Octa}
\end{equation}
Identifying matrices that differ by a multiplicative phase,
we obtain a group of basis transformation with 24 elements, each of which
corresponds to a permutation of the vector labels $abcdef$ as indicated
above.  This group is the projective unitary group $PU(2,9)$
consisting of $2\times 2$ matrices $\underline{U}$ with elements in $GF(9)$ which
satisfy the condition
\begin{equation}
\underline{U}^\dagger \underline{U}\;=\; \pm\mathbf{\1}_{2\times 2}\;,
\end{equation}
with matrices that differ by a multiplicative phase identified.
This group is a subgroup of $PGL(2,9)$ which is isomorphic to
the octahedral group $O$, which is also isomorphic to $S_4$.

\begin{figure}[t]
\sidecaption
\includegraphics[width=7.5cm]{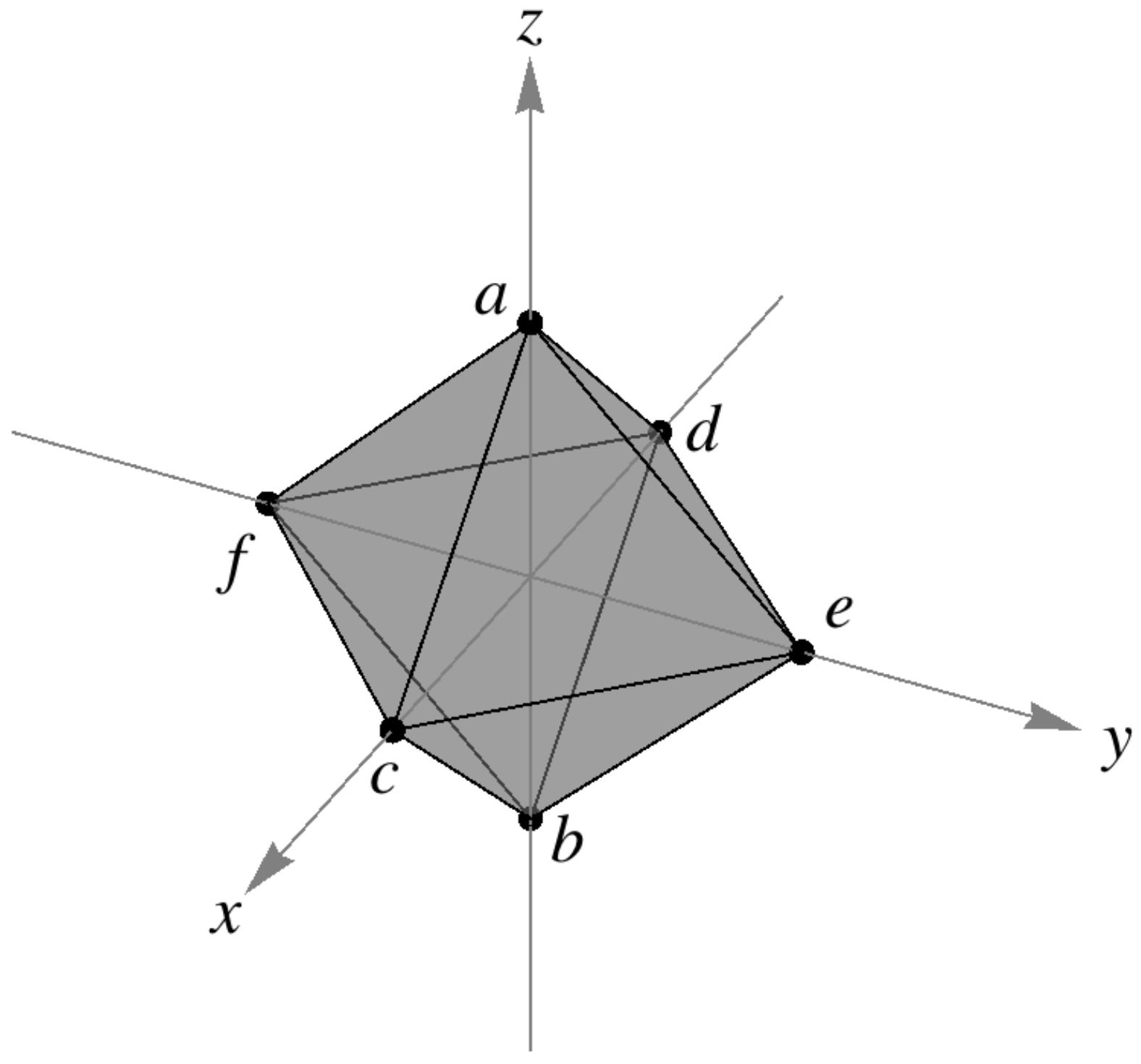}
\caption{The correspondence between the octahedral group $O$ 
and the projective unitary group $PU(2,9)$.
Every $O$ rotation of the octahedron corresponds to a
permutation of the six vertex labels $abcdef$ belonging to $PU(2,9)$.
}
\label{Octafig}
\end{figure}

The isomorphism between $PU(2,9)$ and 
the octahedral group $O$ is implemented by
labeling the six vertices of the octahedron as shown in Fig.~\ref{Octafig}.
Every rotation of the octahedron in $O$ will lead to a permutation of
the vertex labels corresponding to an element of $PU(2,9)$.
For instance, the $180^\circ$ 
rotation around the $x$-axis lead to the permutation $(ab)(ef)$.
The spin observables transform under $PU(2,9)$ as
\begin{eqnarray}
e\; & : & 
\GFsigma_1\rightarrow \GFsigma_1\;,\;
\GFsigma_2\rightarrow \GFsigma_2\;,\;
\GFsigma_3\rightarrow \GFsigma_3\;,
\cr
(ab)(ef) & : & 
\GFsigma_1\rightarrow \GFsigma_1\;,\;
\GFsigma_2\rightarrow -\GFsigma_2\;,\;
\GFsigma_3\rightarrow -\GFsigma_3\;,
\cr
(ab)(cd) & : & 
\GFsigma_1\rightarrow -\GFsigma_1\;,\;
\GFsigma_2\rightarrow \GFsigma_2\;,\;
\GFsigma_3\rightarrow -\GFsigma_3\;,
\cr
(cd)(ef) & : & 
\GFsigma_1\rightarrow -\GFsigma_1\;,\;
\GFsigma_2\rightarrow -\GFsigma_2\;,\;
\GFsigma_3\rightarrow \GFsigma_3\;,
\cr
(aebf) & : & 
\GFsigma_1\rightarrow \GFsigma_1\;,\;
\GFsigma_2\rightarrow -\GFsigma_3\;,\;
\GFsigma_3\rightarrow \GFsigma_2\;,
\cr
(afbe) & : & 
\GFsigma_1\rightarrow \GFsigma_1\;,\;
\GFsigma_2\rightarrow \GFsigma_3\;,\;
\GFsigma_3\rightarrow -\GFsigma_2\;,
\cr
(acbd) & : & 
\GFsigma_1\rightarrow -\GFsigma_3\;,\;
\GFsigma_2\rightarrow \GFsigma_2\;,\;
\GFsigma_3\rightarrow \GFsigma_1\;,
\cr
(adbc) & : & 
\GFsigma_1\rightarrow \GFsigma_3\;,\;
\GFsigma_2\rightarrow \GFsigma_2\;,\;
\GFsigma_3\rightarrow -\GFsigma_1\;,
\cr
(cedf) & : & 
\GFsigma_1\rightarrow \GFsigma_2\;,\;
\GFsigma_2\rightarrow -\GFsigma_1\;,\;
\GFsigma_3\rightarrow \GFsigma_3\;,
\cr
(cfde) & : & 
\GFsigma_1\rightarrow -\GFsigma_2\;,\;
\GFsigma_2\rightarrow \GFsigma_1\;,\;
\GFsigma_3\rightarrow \GFsigma_3\;,
\cr
(ae)(bf)(cd) & : & 
\GFsigma_1\rightarrow -\GFsigma_1\;,\;
\GFsigma_2\rightarrow \GFsigma_3\;,\;
\GFsigma_3\rightarrow \GFsigma_2\;,
\cr
(af)(be)(cd) & : & 
\GFsigma_1\rightarrow -\GFsigma_1\;,\;
\GFsigma_2\rightarrow -\GFsigma_3\;,\;
\GFsigma_3\rightarrow -\GFsigma_2\;,
\cr
(ac)(bd)(ef) & : & 
\GFsigma_1\rightarrow \GFsigma_3\;,\;
\GFsigma_2\rightarrow -\GFsigma_2\;,\;
\GFsigma_3\rightarrow \GFsigma_1\;,
\cr
(ad)(bc)(ef) & : & 
\GFsigma_1\rightarrow -\GFsigma_3\;,\;
\GFsigma_2\rightarrow -\GFsigma_2\;,\;
\GFsigma_3\rightarrow -\GFsigma_1\;,
\cr
(ab)(ce)(df) & : & 
\GFsigma_1\rightarrow \GFsigma_2\;,\;
\GFsigma_2\rightarrow \GFsigma_1\;,\;
\GFsigma_3\rightarrow -\GFsigma_3\;,
\cr
(ab)(cf)(de) & : & 
\GFsigma_1\rightarrow -\GFsigma_2\;,\;
\GFsigma_2\rightarrow -\GFsigma_1\;,\;
\GFsigma_3\rightarrow -\GFsigma_3\;,
\cr
(ace)(bdf) & : & 
\GFsigma_1\rightarrow \GFsigma_2\;,\;
\GFsigma_2\rightarrow \GFsigma_3\;,\;
\GFsigma_3\rightarrow \GFsigma_1\;,
\cr
(adf)(bce) & : & 
\GFsigma_1\rightarrow \GFsigma_2\;,\;
\GFsigma_2\rightarrow -\GFsigma_3\;,\;
\GFsigma_3\rightarrow -\GFsigma_1\;,
\cr
(acf)(bde) & : & 
\GFsigma_1\rightarrow -\GFsigma_2\;,\;
\GFsigma_2\rightarrow -\GFsigma_3\;,\;
\GFsigma_3\rightarrow \GFsigma_1\;,
\cr
(ade)(bcf) & : & 
\GFsigma_1\rightarrow -\GFsigma_2\;,\;
\GFsigma_2\rightarrow \GFsigma_3\;,\;
\GFsigma_3\rightarrow -\GFsigma_1\;,
\cr
(aec)(bfd) & : & 
\GFsigma_1\rightarrow \GFsigma_3\;,\;
\GFsigma_2\rightarrow \GFsigma_1\;,\;
\GFsigma_3\rightarrow \GFsigma_2\;,
\cr
(afc)(bed) & : & 
\GFsigma_1\rightarrow \GFsigma_3\;,\;
\GFsigma_2\rightarrow -\GFsigma_1\;,\;
\GFsigma_3\rightarrow -\GFsigma_2\;,
\cr
(aed)(bfc) & : & 
\GFsigma_1\rightarrow -\GFsigma_3\;,\;
\GFsigma_2\rightarrow -\GFsigma_1\;,\;
\GFsigma_3\rightarrow \GFsigma_2\;,
\cr
(afd)(bed) & : & 
\GFsigma_1\rightarrow -\GFsigma_3\;,\;
\GFsigma_2\rightarrow \GFsigma_1\;,\;
\GFsigma_3\rightarrow -\GFsigma_2\;,
\cr
& &
\end{eqnarray}
just as they should under the corresponding 
rotations of the octahedron in 3D space.

The five conjugacy classes of $PU(2,9)\cong O$ are
\begin{equation}
\begin{array}{l}
\{e\}\;,\\ 
\{(ab)(ef),(ab)(cd),(cd)(ef)\}\;,\\ 
\{(acbd),(adbc),(aebf),(afbe),(cedf),(cfde)\}\;,\\ 
\{(ac)(bd)(ef),(ad)(bc)(ef),(ae)(bf)(cd), \\
\;\;(af)(be)(cd),(ab)(ce)(df),(ab)(cf)(de)\}\;,\; \mbox{and} \\
\{(ace)(bdf),(adf)(bce),(acf)(bde),(ade)(bcf),\\
\;\;(aec)(bed),(add)(bed),(add)(bfc),(afc)(bed)\}\;.\\
\end{array}
\label{Octaclasses}
\end{equation}
The 504 physical entangled states in $\V{2}{9}\times\V{2}{9}=\V{4}{9}$
also fall into classes that transform among themselves under global
$PU(2,9)$ transformations.
Since we cannot list all 504 states here, we will only mention that
they fall into 17 classes of 24 elements each, 4 classes of 12 elements each,
4 classes of 8 elements each, 2 classes of 6 elements each, 1 class of 3 elements,
and the singlet state $\ket{S}=[\,\0,\1,-\1,\0\,]^\mathrm{T}$. 

This can be verified by a direct search, or through use of Burnside's lemma and 
the orbit-stabilizer theorem. For a group, $G$, acting on a set, $X$, a subset that is preserved by the 
action of the entire group is called an orbit. The set of orbits forms a partition of the set $X$. 
To calculate the number of orbits, here denoted $|X/G|$, one can use Burnside's lemma:
\begin{equation}
|X/G| \;=\; \frac{1}{|G|}\sum_{g\in G}|X^g|\;,
\end{equation} 
where $X^g$ is the set of elements in $X$ that are invariant under the action of $g$.

For global $PU(2,9)$ transformations, Burnside's lemma indicates that there should be 
29 orbits in the set of entangled states. To calculate the length of these orbits, 
one could use the orbit-stabilizer theorem, which states that the order of the orbit containing an 
element is equal to the order of the group divided by the order of the stabilizer 
subgroup of that element. The stabilizer subgroup of an element is the subgroup under 
which that element is invariant.

This computation indicates that there are 408 states that belong to orbits of order 24, 
48 states that belong to orbits of order 12, 32 states that belong to orbits of order 8, 
12 states that belong to orbits of order 6, 3 states that belong to orbits of order 3, and
1 states that belongs to an orbit of order 1.

Under local $PU(2,9)$ transformations, the same 504 entangled states fall into
three classes with 24, 288, and 192 elements each. Again, this result can be arrived at 
through a manual search or through the group theoretic means mentioned above.
Representative elements from the three classes can be taken to be:
\begin{eqnarray}
\ket{S} & = & \bigl[\begin{array}{cccc} \0 & \1 & -\1 & \0 \end{array}\bigr]^\mathrm{T}\;,\cr
\ket{T} & = & \bigl[\begin{array}{cccc} \1 & \0 & \1+\i & \1 \end{array}\bigr]^\mathrm{T}\;,\cr
\ket{U} & = & \bigl[\begin{array}{cccc} \1 & \0 & \1 & \1+\i \end{array}\bigr]^\mathrm{T}\;.
\end{eqnarray}
Therefore, to obtain the CHSH bound, we only need to calculate the correlators
for these three states.

\section{Uniqueness of the Product Preserving Map}

Is the $\ppm$ function defined in Eq.~(\ref{phidef}) the only function that allows us to 
calculate real expectation values? Yes, as we will see via the following argument.

What are the physical requirements on $\ppm$? Primarily, it must be a map from 
$GF(p)$ to $\R$, as we assume that the results of measurements 
are real numbers. For the expectation value of the identity operator to be 1, we must have 
that $\ppm(\1)=1$. Likewise, for the expectation value of the zero operator to be 0, 
we must have that $\ppm(\0)=0$. When considering two particle states, if we require that the expectation 
values of product states should factorize, then $\ppm$ must respect multiplication. This is done if the image of $GF(p)\backslash\{\0\}$ is homomorphic to $GF(p)\backslash\{\0\}$. Since $GF(p)\backslash\{\0\}$ is cyclic,
any group homomorphic to it must also by cyclic. The only cyclic, multiplicative subgroups of $\R$ are $\{+1\}$, $\{0\}$, and 
$\{+1,-1\}$.  

As the image must contain $+1$, $\{0\}$ is excluded. If we choose $\{+1,-1\}$ as the image of 
$GF(p)\backslash\{\0\}$ under $\ppm$, we are allowed to interpret eigenvalues of observables as the 
expectation values of the corresponding eigenstates. Thus, the image of $GF(p)\backslash\{\0\}$ 
should be $\{+1,-1\}$ and $\ppm$ should be surjective between $GF(p)\backslash\{\0\}$ and $\{+1,-1\}$, 
to ensure the presence of $-1$ as an expectation value.  

For $\ppm$ to be a such a surjection, it must be true that $\ppm(\g)=-1$ whenever $\g$ is a 
multiplicative generator of $GF(p)\backslash\{\0\}$.  It then follows that all even powers of $\g$ should map to $+1$.
Thus, the kernel of $\ppm$ must contain all $(p-1)/2$ even powers of $\g$. Note that it does not matter which generator is chosen since any given generator is an odd power of each of the other generators.

The kernel of a group homomorphism, the set of elements that map to the identity, is a subgroup. In order for 
$\ppm$ to be surjective from $GF(p)\backslash\{\0\}$ to $\{+1,-1\}$, its kernel must be a proper subgroup of $GF(p)\backslash\{\0\}$. As we have shown that 
the kernel must contain half of the elements of $GF(p)\backslash\{\0\}$, it can only contain those elements, as the 
order of the kernel must divide the order of $GF(p)\backslash\{\0\}$. Therefore, $\ppm$ as defined in Eq.~(\ref{phidef}) is the only map that fits the relevant criteria.


\end{document}